\documentclass[twocolumn]{aastex631}

\usepackage{multirow}
\usepackage{natbib}

\defcitealias{2021NatAs...5..928S}{S21}
\defcitealias{2022ApJ...926..209S}{S22}

\shorttitle{AASTeX v6.3.1 Sample article}
\shortauthors{Shi et al.}
\graphicspath{{./}{figures/}}

\begin{document}

\title{Observational Evidence for Hot Wind Impact on pc-scale in Low-luminosity Active Galactic Nucleus}

\author[0000-0003-3922-5007]{Fangzheng Shi}
\affiliation{Shanghai Astronomical Observatory, Chinese Academy of Sciences, Shanghai, People's Republic of China}\email{fzshi@shao.ac.cn}

\author[0000-0003-3564-6437]{Feng Yuan}
\affiliation{Center for Astronomy and Astrophysics and Department of Physics, Fudan University, Shanghai 200438,
People's Republic of China}\email{fyuan@fudan.edu.cn}

\author[0000-0003-0355-6437]{Zhiyuan Li}
\affiliation{School of Astronomy and Space Science, Nanjing University, Nanjing, People's Republic of China}\email{zyli@@nju.edu.cn}

\author[0000-0002-6738-3259]{Zhao Su}
\affiliation{School of Astronomy and Space Science, Nanjing University, Nanjing, People's Republic of China}

\author[0000-0001-9658-0588 ]{Suoqing Ji}
\affiliation{Shanghai Astronomical Observatory, Chinese Academy of Sciences, Shanghai, People's Republic of China}

\begin{abstract}
Supermassive black holes in galaxies spend majority of their lifetime in the low-luminosity regime, powered by hot accretion flow. Strong winds launched from the hot accretion flow have the potential to play an important role in active galactic nuclei (AGN) feedback. Direct observational evidence for these hot winds with temperature around 10 keV, has been obtained through the detection of highly ionized iron emission lines with Doppler shifts in two prototypical low-luminosity AGNs, namely M81* and NGC\,7213.
In this work, we further identify blueshifted H-like O/Ne emission lines in the soft X-ray spectra of these two sources. These lines are interpreted to be associated with additional outflowing components possessing velocity around several $10^3$ km/s and lower temperature ($\sim 0.2-0.4$ keV).
Blue-shifted velocity and the X-ray intensity of these additional outflowing components are hard to be explained by previously detected hot wind freely propagating to larger radii.
Through detailed numerical simulations, we find the newly detected blue-shifted emission lines would come from circumnuclear gas shock-heated by the hot wind instead.
Hot wind can provide larger ram pressure force on the clumpy circumnuclear gas than the gravitational force from central black hole, effectively impeding the black hole accretion of gas.
Our results provide strong evidences for the energy and momentum feedback by the hot AGN wind. 

\end{abstract}


\section{Introduction} \label{sec:intro}

Actively accreting supermassive black holes (SMBHs) located at the center of galaxies can considerably influence the evolution of their host galaxies through radiation or different kinds of outflows \citep[e.g.,][]{2013ARA&A..51..511K}. Accretion and corresponding feedback can be classified into hot and cold modes, depending on whether the mass accretion rate of the SMBHs is below or above $\sim 2\%$ of $\dot{M}_{\rm Edd}$ \citep{2014ARA&A..52..529Y}. The majority of local galaxies host low-luminosity active galactic nuclei (LLAGNs), resulting in hot (kinetic) mode accretion and feedback within these galaxies.

Numerical simulations of hot accretion flows have predicted the existence of energetic outflows in the form of uncollimated hot winds with a large opening angle \citep{2012ApJ...761..130Y,2012ApJ...761..129Y,2012MNRAS.426.3241N,2015ApJ...804..101Y,2021ApJ...914..131Y}.
Furthermore, simulations have shown that, unless the black hole spin is extremely high, the kinetic energy and momentum carried by the hot wind will be comparable, if not larger than that carried by jet \citep{2015ApJ...804..101Y,2021ApJ...914..131Y}.
This result, combined with the fact that the opening angle of wind is much larger than that of jet, suggests that hot winds serve as an essential component of kinetic mode feedback, in addition to the widely considered relativistic jets. The importance of the hot winds and radiation in the hot feedback mode has been investigated by numerical simulations of AGN feedback in an isolated elliptical galaxy \citep{2019ApJ...885...16Y}. Adopting {\it MACER} model \citep{2018ApJ...857..121Y}, where radiation and wind of AGN have been carefully incorporated based on the state-pf-the-art physics of black hole accretion in both the hot (radio) and cold (quasar) modes,
they found AGN activity and black hole mass would become significantly higher if hot mode feedback is turned off. And the suppression effect of hot mode feedback is dominated by wind rather than radiation.

Direct observational evidence for the existence of hot winds has been reported in two LLAGNs, namely M81* and NGC\,7213, through Doppler shift of highly ionized Fe emission lines from observations conducted by Chandra High Energy Transmission Gratings (HETG) by \citet{2021NatAs...5..928S,2022ApJ...926..209S} (hereafter \citetalias{2021NatAs...5..928S,2022ApJ...926..209S}).
M81* has an Eddington ratio of is $L_{\rm bol}/L_{\rm Edd}\sim3\times10^{-5}$ \citep{2011ApJ...726...87Y,2014MNRAS.438.2804N}, well below the threshold for LLAGN ($\sim2\%$) \citep{2014ARA&A..52..529Y}.
Symmetrically red- and blue-shifted Fe XXVI Ly$\alpha$ emission line with consistent velocity to the rest frame energy have been detected, indicating the presence of a 16-keV bipolar outflow with line-of-sight velocity of $\pm2800$ km/s.
The Eddington ratio of NGC\,7213 is $10^{-3}$ \citep{2005MNRAS.356..727S,2012MNRAS.424.1327E}. Highly ionized iron lines, including Fe XXV K$\alpha$ and Fe XXVI Ly$\alpha$ lines, have been detected \citep{2003A&A...407L..21B,2005MNRAS.356..727S,2013MNRAS.429.3439E} and appear to have a blueshifted velocity of $\sim1000$ km/s \citep{2008MNRAS.389L..52B,2022ApJ...926..209S}. 
These blueshifted emission lines are interpreted by a 16-keV hot wind with a velocity of 1100 km/s from the hot accretion flows \citepalias{2022ApJ...926..209S}. 

The kinetic luminosity carried by hot winds accounts for 10\% - 15\% of the AGN bolometric luminosity in M81* and NGC\,7213. 
However the spatial extent of the hot winds influence remains uncertain.
In both M81* and NGC\,7213, static hot gas with $T\simeq1$-keV, potentially associated with hot circumnuclear medium, is spatially confined to the central $10^2 - 10^3$ pc \citepalias{2021NatAs...5..928S,2022ApJ...926..209S}, implying that such hot winds may have the capability to affect the evolution of circumnuclear medium.
The presence of multiphase outflows at several $10^2$ pc-scale is hinted by multiwavelength observations in these two sources.
Biconical warm ionized gas outflow features with a line-of-sight velocity of $\sim50$ km/s and total mass outflow rate of $\sim1.4\times10^{-3}\rm~M_{\odot}~yr^{-1}$ have been detected within 120 - 250 pc of M81* \citep{2022ApJ...928..111L}.
The mass outflow rate is similar to that of M81* pc-scale hot wind ($\sim2\times10^{-3}\rm~M_{\odot}~yr^{-1}$).
In the case of NGC\,7213, two potential molecular outflows within 500 pc of the central SMBH have been detected \citep{2020A&A...641A.151S}.
The connection between these larger-scale outflows and pc-scale hot winds requires further investigation.

The large-scale dynamics of wind after launching from accretion flow has been studied by \citet{2020ApJ...890...80C} and \citet{2020ApJ...890...81C} \citepalias[see also][]{2021NatAs...5..928S,2022ApJ...926..209S}. These studies predict that, without considering the interaction of the wind with interstellar medium (ISM), the wind temperature should gradually decrease from $10^{8-9}$ K to $10^6$ K at $\gtrsim10^6\rm~r_{g}$, while the wind velocity remains almost unchanged or slightly increases, depending on the strength and configuration of magnetic field.
As a result, we would expect to observe blueshifted thermal emission lines of highly ionized O and Ne from $\sim10^6$K plasma as the winds propagate to outer regions.
Observations with high spatial and spectral resolution in the soft X-ray regime are necessary to explore the impact of hot winds on host galaxies, which is the aim of this work.

This paper is organized as follows. In Section \ref{sec:data}, we briefly describe the reduction process of X-ray observations involved in this study.
We analyze the high-resolution X-ray spectra of M81 and NGC\,7213, focusing on the detection of prominent emission lines in soft X-ray bands, and discuss potential contamination in Section \ref{sec:spec}.
In Section \ref{sec:diss} we associate the identified blue-shifted emission lines with bulk motion plasma. We address the challenges faced by free-expanding wind model in describing this outflowing plasma. And we conclude the observed spectral features can be better explained by winds interact with circumnuclear warm ionized medium through numerical simulations.
We summarize our results in Section \ref{sec:sum}.
The errors of derived parameters listed in this work are at 90\% confidence level unless otherwise stated.

\section{Data Reduction} \label{sec:data}
The XMM-Newton Reflection Grating Spectrometer (RGS) can achieve a full-width half-minimum (FWHM) spectral resolution of $\sim1200$ km/s for the 1st-order over 0.5-2 keV soft X-ray bands, which meets the requirements for detecting putative hot winds with several $10^3\rm~km/s$. 
M81* has been observed by XMM-Newton with RGS on 2001 April 23th for 132.7 ks ({\it ObsID: 0111800101}). There is a total RGS 182.1 ks exposure of NGC\,7213 on 2001 May 28 and 2009 Nov 11 ({\it ObsID: 0111810101 and 0605800301}). 

The RGS data are reduced with the XMM-Newton Science Analysis System (SAS)\footnote{\url{http://www.cosmos.esa.int/web/XMM-Newton/sas}} v20.0.0.
We follow the standard pipeline {\it rgsproc} to clean and reprocess the observations.
After removing bad time interval with flaring particle background, we obtain a clean exposure time of 87.0 ks for M81* and 159.2 ks for NGC\,7213.
To optimize the signal-to-noise ratio, we test different widths of source extraction regions along the cross-dispersion direction to include 60\% - 95\% of line spread function (LSF).
The background spectra are extracted from regions outside 98\% of LSF.
Ultimately, a source extraction region including 90\% of LSF ($\sim1.46\arcmin$) is chosen.
We combine the 1-order spectra from the two RGS spectrometer units (Fig. \ref{fig:BLsearch}).
Spectra are regrouped to ensure that each energy bin contains at least one count.

\section{Spectral Analysis} \label{sec:spec}
\subsection{Baseline Continuum Modeling} \label{subsec:baseline}
Spectra for the two sources are analyzed with Xspec v12.12.1 and Cash statistics ($C$-stat) is adopted to evaluate the goodness of fit. 
Galactic foreground absorption is always included in the following fitting procedure. The absorption column density is fixed at $N_{\rm H}=1\times10^{21}\rm~cm^{-3}$ for M81* and $N_{\rm H}=1.08\times10^{20}\rm~cm^{-3}$ for NGC\,7213\footnote{Values of galactic absorption column density are obtained from \url{https://heasarc.gsfc.nasa.gov/cgi-bin/Tools/w3nh/w3nh.pl}.}. 
To fit the continuum of stacked spectrum for each source, a simple power-law model is initially adopted.
Figure \ref{fig:BLsearch} shows the fitting results of the baseline continuum models and Table \ref{tab:continuum} lists the derived parameters, which will be used for line diagnostics in the following sections.

\begin{deluxetable*}{c|l|ccccccccc|c}[htbp]
\tabletypesize{\scriptsize}
\tablecaption{Continuum Fitting Results\label{tab:continuum}}
\tablehead{
\colhead{Source} & \colhead{model} & \colhead{$\Gamma$} & \colhead{$N_{\rm pl}$} &
\colhead{$T_0$} & \colhead{$N_0$} & \colhead{$T_{\rm b}$} & 
\colhead{$v_{\rm b}$} & \colhead{$A_{\rm O}$} & \colhead{$A_{\rm Ne}$} & \colhead{$N_{\rm b}$} & \colhead{$C$-stat} \\
\colhead{(1)} & \colhead{(2)} & \colhead{(3)} & \colhead{(4)} &
\colhead{(5)} & \colhead{(6)} & \colhead{(7)} & \colhead{(8)} &
\colhead{(9)} & \colhead{(10)} & \colhead{(11)} & \colhead{(12)}
}
\startdata
\multirow{3}{*}{M81} & PL (order 1) & $2.19\pm0.03$ & $4.44\pm0.04$ & - & - & - & - & - & - & - & 1904/1827\\
{   } & PL+AP$_0$ & 2.19 & $4.22\pm0.07$ & $0.58_{-0.08}^{+0.09}$ & $37_{-8}^{+9}$ & - & - & 1 & $2.1_{-0.8}^{+1.0}$ & - & 1818/1825\\
{   } & PL+AP$_0$+AP$_b$ & 2.19 & $4.17\pm0.06$ & $0.57\pm0.09$ & $32\pm7$ & $0.44_{-0.09}^{+0.13}$ & $1.9^{+0.4}_{-0.3}$ & $0.4\pm0.2$ & $2.1_{-0.7}^{+1.8}$ & $16\pm7$ & 1807/1821\\
\hline
\multirow{3}{*}{NGC\,7213} & PL (order 1) & $1.88\pm0.02$ & $4.10\pm0.03$ & - & - & - & - & - & - & - & 2152/1834 \\
{         } & PL+AP$_0$ & 1.88 & $4.05\pm0.03$ & $0.27^{+0.05}_{-0.04}$ & $6.4\pm1.7$ & - & - & - & - & - & 2109/1833\\
{         } & PL+AP$_0$+AP$_b$ & 1.88 & $4.05\pm0.03$ & 0.27 & $6.5_{-1.8}^{+1.6}$ & $0.20^{+0.07}_{-0.05}$ & $2.7\pm0.3$ & 1 & 1 & $3.4_{-1.9}^{+2.0}$  &2101/1831\\
\enddata
\tablecomments{
(1) Source name. 
(2) Model combinations of the spectral fitting. 'PL' stands for power-law and 'AP' represents the {\it apec} model. Subscript '0' and 'b' denote the static and outflowing components. 
(3) Power-law photon index. 
(4) Normalization of the power-law, in units of $\rm 10^{-3}~ph~s^{-1}~cm^{-2}~keV^{-1}$ at 1 keV. 
(5) \& (7) Plasma temperature of static and blue-shifted gas components. 
(6) \& (11) Normalization of the {\it apec} components, in units of $\rm 10^{-5}~cm^{-5}$.
(8) Blue-shifted velocity in units of $\rm 10^{3}$ km/s.
(9) - (10) Abundance of Ne and O relative to solar abundance.
(12) Goodness of fit in {\it C}-stat and degree of freedom.
}
\vspace{-4em}
\end{deluxetable*}

For M81, the best-fit photon index $\Gamma=2.19\pm0.03$ with $C$-stat=1901 for 1827 degree of freedom (d.o.f), which is softer than the photon index derived with Chandra HETG and NuSTAR spectra in 1-79 keV band in our previous work \citepalias{2021NatAs...5..928S}, probably indicate the
existence of additional thermal gas components with temperature less than 1 keV. This component would contribute to soft band and makes the power-law component steeper.

For NGC\,7213, the best-fit photon index is $\Gamma=1.877\pm0.019$ with $C$-stat=2152 for 1834 d.o.f.. The photon index is also a little softer compared with the joint fit results from Chandra HETG and NuSTAR FPMA/FPMB\citepalias{2022ApJ...926..209S}.
We also check the spectral variability between two RGS epochs of NGC\,7213 spectra. The best-fit photon index is marginally consistent cross epochs($1.88\pm0.03$ and $1.84\pm0.03$ respectively). While photon flux declines $\sim$45\% over the 8-year interval, which is in agreement with the descending trend described by \citet{2018MNRAS.475.1190Y}. 
Emission line residues can be clearly seen in both the co-added and individual epoch spectra.

\subsection{Line diagnostic} \label{subsec:blsearch}

\begin{figure*}
\centering
\includegraphics[width=0.95\linewidth]{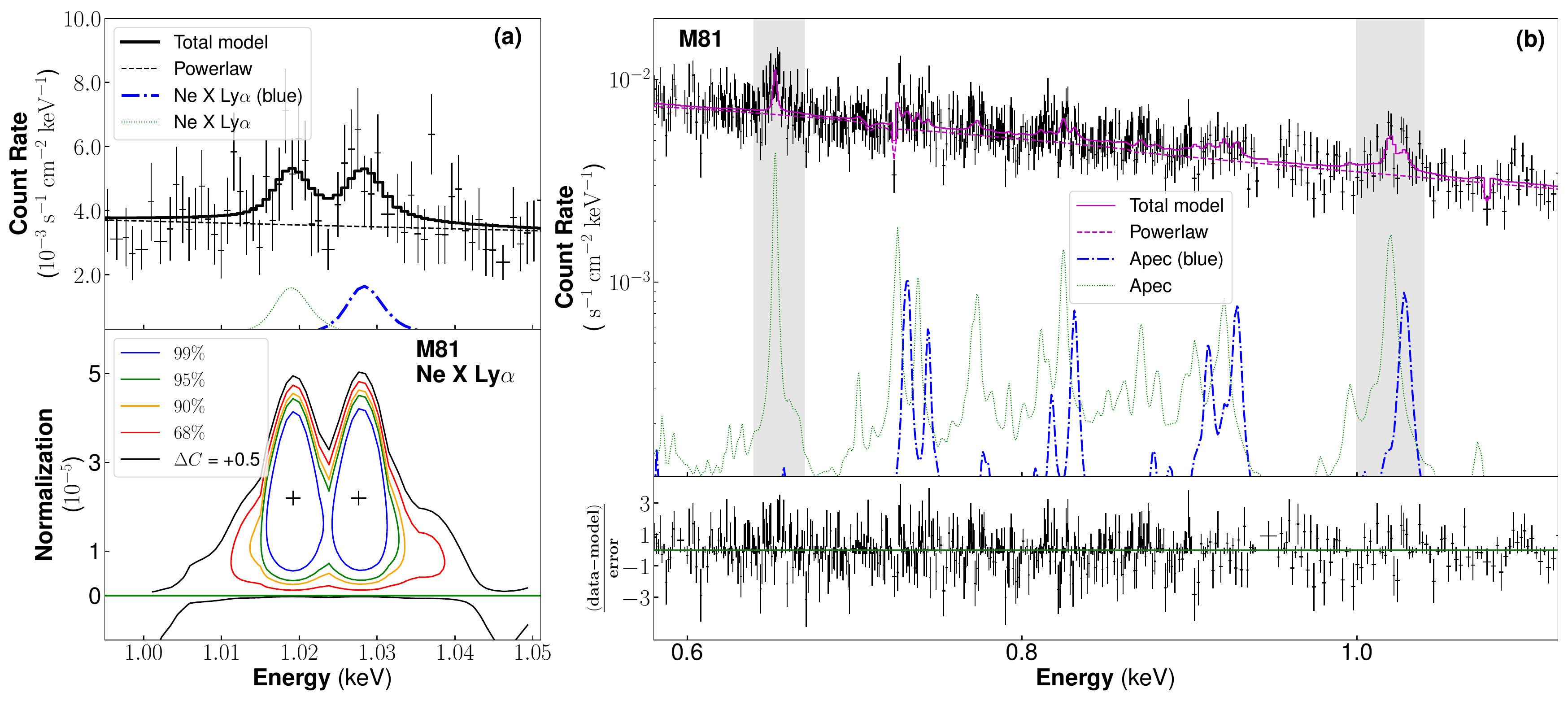}
\includegraphics[width=0.95\linewidth]{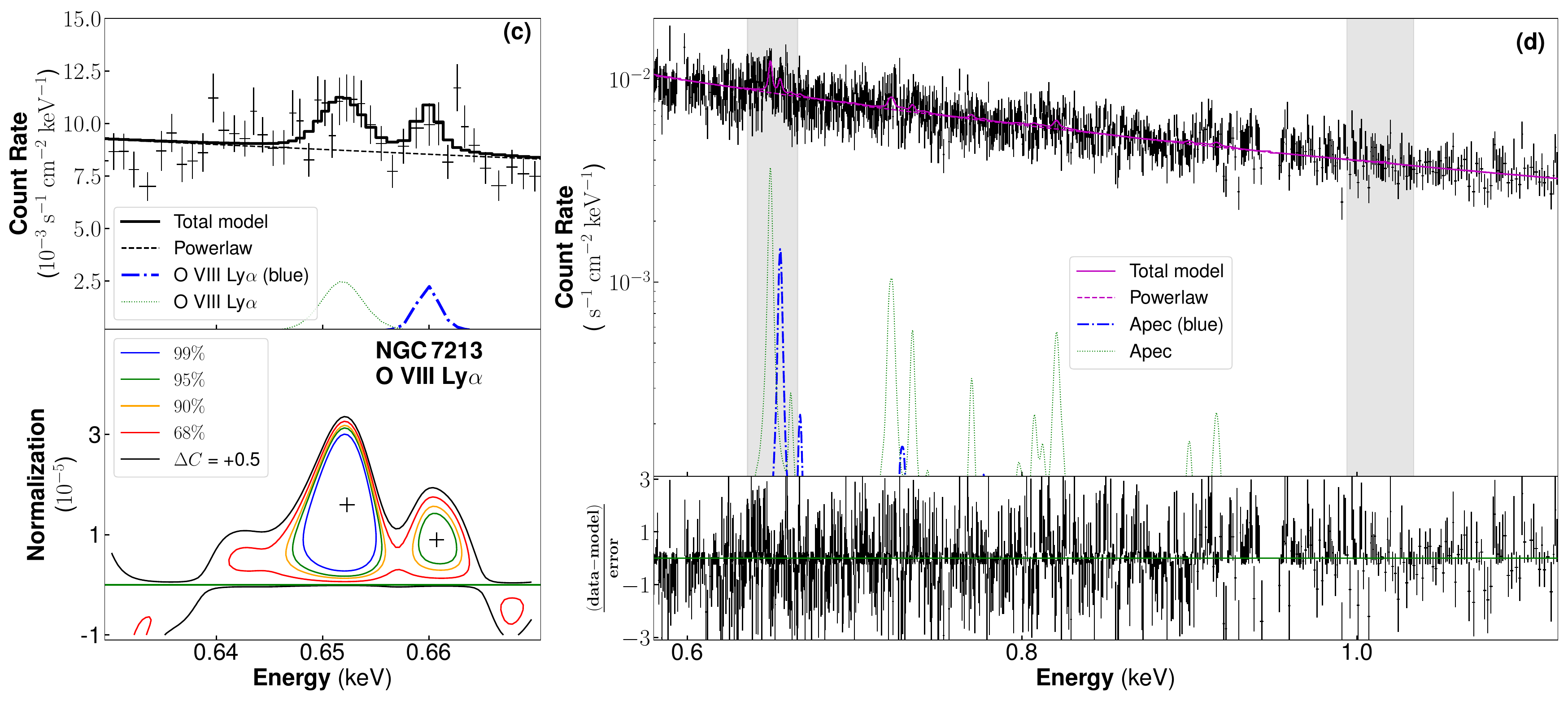}
\caption{
{\bf (a) \& (c)}:
Gaussian fitting (up) and confidence level contours (bottom) of extra prominent blue-shifted atomic transition lines (blue dash-dotted lines) paired with rest-frame counterparts (green dotted lines), Ne X Ly$\alpha$ for M81* and O VIII Ly$\alpha$ for NGC\,7213 in the RGS spectra.
Black solid and dashed lines describe the total and power-law continuum models.
Blue, green, yellow and red contours correspond to confidence level of 99\%, 95\%, 90\%, 68\% respectively, which is derived from the improvement of $C$-stat for an additional Gaussian line with respect to the baseline continuum model.
The photon energy has been calibrated for systematic redshift. 
{\bf (b) \& (d)}: 
Best-fit spectra and the ratio between residues and errors over 0.6-1.1 keV band for M81* and NGC\,7213. The blue-shifted components (blue dashed-dot lines) are modeled with single-temperature, single-velocity thermal emission model {\it apec}. 
The additional static thermal components {\it apec} are represented in green dotted lines.
Grey stripes mark the energy bands where highly ionized O and Ne emission lines reside.
Black crosses mark the observed spectral data points with 1$\sigma$ error bar. 
The spectra are optimally binned to ensure signal-to-ratio\textgreater3 for illustration purpose.}
\label{fig:BLsearch}
\end{figure*}

We perform a blind line search for the RGS spectrum of each source to identify prominent emission lines in soft X-ray band.
Systematic redshift has been corrected.
Detailed description of blind line search method can be found in \citet{2010A&A...521A..57T} and \citetalias{2021NatAs...5..928S}. 
We adopt the Monte Carlo Markov Chain script embedded in {\it Xspec} to check and constrain the 90\% error range of the line centroids.
Identification and information of recognized lines are summarized in Table \ref{tab:emisLine} with the help of ATOMDB\footnote{\url{http://atomdb.org/}\label{atomdb}}.
Map of confidence level map for detected prominent emission lines are shown in Figure \ref{fig:BLsearch}.

In the soft X-ray band spectrum of M81, we detect strong O VIII Ly$\alpha$ emission line at $0.6534_{-0.0010}^{+0.0016}$ keV, which is slightly resolved with $\sigma=9_{-7}^{+8}\times10^2\rm~km~s^{-1}$. 
Emission line centered at $0.562_{-0.007}^{+0.003}$ keV with 3-$\sigma$ upper limit line width of \textless $4.3\times10^3\rm~km~s^{-1}$ is assigned to the forbidden line of He-like O VII K$\alpha$ triplet. Corresponding resonant and inter-combination counterparts can be marginally detected.
In the vicinity of the rest-frame energy of Ne X Ly$\alpha$ line (1.022 keV), two distinct unresolved emission lines are detected with at least 3-$\sigma$ confidence level. One of the lines centered at $1.019_{-0.001}^{+0.002}$ keV and is consistent with the rest-frame H-like Ne X line. The other line, lying at $1.028\pm0.002$ keV, exhibits blue-shifted velocity of $1.9^{+0.6}_{-0.4}\times10^3\rm~km~s^{-1}$ with respect to rest-frame Ne X Ly$\alpha$ centroid.
This blue-shifted Ne X Ly$\alpha$ line is also hinted in the blind line search result of co-added Chandra HEG spectrum\citepalias{2021NatAs...5..928S}, centered at $1.026_{-0.003}^{+0.004}$ keV with confidence level of 98.3\%. The line flux is consistent with 3-sigma upper limit in MEG spectrum, but several times smaller than that in RGS.

In NGC\,7213, we  two marginally resolved emission lines centered on $0.563_{-0.007}^{+0.002}$ keV and $0.574\pm0.04$ keV, which could be associated with the forbidden and resonance component He-like O VII K$\alpha$ line respectively.
Emission lines around $0.727_{-0.005}^{+0.002}$ keV and $0.741\pm0.005$ keV are assigned to Fe XVII K$\alpha$ lines.
Ne IX resonance line is detected at $0.917^{+0.005}_{-0.004}$ keV.
There is no obvious residue around the rest-frame energy of Ne X Ly$\alpha$ line. 
We detect $0.652\pm0.002$ keV O VIII Ly$\alpha$ line with a line width of $6_{-3}^{+3}\times10^2$ km/s. Besides, an additional emission line has been detected at 98.4\% confidence level $3^{+3}_{-2}\times10^3$ km/s blue-shifted with respect to the rest-frame energy of O VIII Ly$\alpha$ line.

\subsection{Spectral modeling}
\label{subsec:spec_mod}
Detected emission lines can be produced by optically thin hot gas in the vicinity of the central AGN. Lines with Doppler shifts imply gas with bulk motion.
We adopt a thermal plasma model {\it apec} (denoted as `AP$\rm_{0}$') corrected for systematic redshift to describe the static gas component and an additional {\it apec} model (denoted as `AP$\rm_{b}$') with Doppler shift $z$ free to vary accounting for the outflowing hot plasma.
Table \ref{tab:continuum} records the best-fit parameters of the gas components.
Contamination from off-nucleus point sources has been discussed in Appendix \ref{appendix:contam} as well as Fig. \ref{fig:off_nu}, and would not cause significant changes to the best-fit values.

The best fit temperature of static gas component in NGC\,7213 is $0.27_{-0.04}^{+0.05}$ keV, with a total 0.5-2 keV luminosity of $4.3^{+1.2}_{-1.1}\times10^{39}$ erg/s.
Inclusion of a blue-shifted thermal component with best-fit temperature $0.20^{+0.07}_{-0.05}$ and velocity of $2.7\pm0.3\times10^3$ km/s would bring an improvement of 8.5 for extra two degree of freedom to the goodness of fit, with 0.5-8 keV luminosity of $2.0^{+1.1}_{-1.0}\times10^{39}\rm~erg~s^{-1}$.

For M81*, the case is a little complicated. 
The temperature of the almost static circumnuclear hot gas around M81 is $0.57\pm0.09$ keV with $L_{\rm 0.5-2}=3.3^{+0.6}_{-0.6}\times10^{38}$ erg/s. This component manifests marginal red-shift with 3-$\sigma$ upper limit of $\lesssim450$ km/s.
Adding another blue-shifted {\it apec} component with bulk motion velocity of $1.9^{+0.4}_{-0.3}\times10^3$ km/s will bring an improvement of 11 for 4 d.o.f in {\it C}-stat.
The temperature of this blue-shifted component is $0.44^{+0.13}_{-0.09}$ keV and its 0.5-8 keV luminosity is estimated to be $2.5_{-1.0}^{+1.1}\times10^{38}$ erg/s.
To achieve this fit, Ne abundance should be super-solar ($A_{\rm Ne}\sim2.1\pm0.5$) in both static and the fast moving gas, and O abundance would be sub-solar ($A_{\rm O}\sim0.4\pm0.2$) in the blue-shifted gas relative to solar abundance taken from taken from \citet{2009ARA&A..47..481A}.
This is dominated by the enhancement in Ne X Ly$\alpha$ and suppression in O VIII Ly$\alpha$ lines, probably due to the self-absorption of certain bound-bound transition in hot winds, which has been discussed in detail in Sec. \ref{subsec:freeWind} and Appendix \ref{appendix:opdepth}.

\begin{deluxetable*}{c|lcccccc}[!ptb]
\tablecaption{Significant Emission Lines\label{tab:emisLine}}
\tablehead{
\colhead{Source} & \colhead{Line} & \colhead{$E_{\rm lab}$} & \colhead{$E_{\rm bl}$} &
\colhead{Significance} & \colhead{$E_{\rm g}$} & \colhead{$\sigma_{\rm g}$} & \colhead{F}\\
\colhead{(1)} & \colhead{(2)} & \colhead{(3)} & \colhead{(4)} & \colhead{(5)} & \colhead{(6)} &
\colhead{(7)} & \colhead{(8)}
}
\startdata
\multirow{5}{*}{M81} & O VII K$\alpha$ (f) & 0.5611 & 0.5619 & 93.61\% & $0.5619_{-0.007}^{+0.003}$ & \textless 8 & $1.8_{-1.2}^{+1.3}$ \\
& O VIII Ly$\alpha$ & 0.6536 & 0.6541 & \textgreater 99.99\% & $0.6534_{-0.0010}^{+0.0016}$ & $2_{-1.5}^{+1.7}$ & $2.8_{-0.9}^{+1.0}$ \\
& Ne X Ly$\alpha$ & 1.0218 & 1.0192 & 99.99\% & $1.019_{-0.001}^{+0.002}$ & 0 & $3.1_{-1.5}^{+1.6}$\\
& Ne X Ly$\alpha$ (blue) & 1.0218 & 1.0276 & 99.99\% & $1.028\pm0.002$ & 0 & $3.2_{-1.5}^{+1.6}$\\
\hline
\multirow{7}{*}{NGC\,7213} & O VII K$\alpha$ (f) & 0.5611 & 0.5629 & 99.91\% & $0.563_{-0.007}^{+0.002}$ & $0.6\pm0.2$ & $0.8_{-0.6}^{+0.7}$ \\
& O VII K$\alpha$ (r) & 0.5740 & 0.5736 & 98.14\% & $0.574\pm0.04$ & \textless 4 & $1.0_{-0.6}^{+0.7}$ \\
& O VIII Ly$\alpha$ & 0.6536 & 0.6520 & \textgreater 99.99\% & $0.652\pm0.002$ & $1.4\pm0.6$ & $1.6_{-0.5}^{+0.6}$ \\
& O VIII Ly$\alpha$ (blue) & 0.6536 & 0.6607 & 98.40\% & $0.660\pm0.005$ & \textless 5 & $0.7\pm0.5$\\
& Fe XVII K$\alpha $ (2-1/3-1) & 0.7263 & 0.7278 & 97.31\% & $0.727_{-0.005}^{+0.002}$ & $1.4^{+3}_{-1.1}$ & $0.8_{-0.5}^{+0.6}$ \\
& Fe XVII K$\alpha $ (5-1) & 0.7389 & 0.7404 & 99.32\% & $0.741\pm0.005$ & \textless 20 & $0.9\pm0.5$ \\
& Ne IX K$\alpha$ (r) & 0.9220 & 0.9174 & 99.38\% & $0.917^{+0.005}_{-0.004}$ & \textless 4 & $1.4_{-0.7}^{+0.9}$ \\
\enddata
\tablecomments{
(1) Source name.
(2) - (3) Identified line transition labels and line centroid in laboratory frame in units of keV.
(4) Most probable line centroid derived from blind line search with systematic redshift corrected, in units of keV.
(5) Confidence level of line detection.
(6) Line centroid determined by Gaussian model fitting, in units of keV.
(7) Line width in units of eV.
(8) Line flux in units of $\rm10^{-14}~erg~s^{-1}~cm^{-2}$.
}
\vspace{-3em}
\end{deluxetable*}

\section{Discussion} \label{sec:diss}

From the above spectral analysis, several H-like and He-like emission lines consistent with rest-frame line centroids are detected with at least 95\% confidence level in soft X-ray bands of M81* and NGC\,7213, associated with a static thermal gas component. 
While an additional outflowing plasma component is traced by blue-shifted Ne X Ly$\alpha$ in M81* and O VIII Ly$\alpha$ lines in NGC\,7213.
In this section we discuss several possible origins of these components.

\subsection{Free-expanding wind} \label{subsec:freeWind}

The outflowing plasma with lower temperature of $\sim0.2-0.4$ keV can be associated with wind expanding to larger radii and experiencing a decrease in temperature.

To roughly estimate the strongest emitting location $r_{\rm aver}$ of wind components detected in \citetalias{2021NatAs...5..928S}, \citetalias{2022ApJ...926..209S} and this work, 
a theoretical investigation of radial temperature and density profile should be established.
In the previous work of \citetalias{2021NatAs...5..928S} and \citetalias{2022ApJ...926..209S}, customized magneto-hydrodynamic simulations have been conducted for M81* (NGC\,7213) to study the propagation of hot winds launched from hot accretion flow without considering the interaction with ISM.
In the simulations, central black hole mass is set to be $7\times10^{7}\rm~M_{\odot}$ ($10^8\rm~M_{\odot}$). The inner boundary of the simulation domain is set to the truncation at radius 3000 (2000) $\rm r_g$, which is the outer boundary of the hot accretion flow, consistent with that derived from Fe I K$\alpha$ line and reflection fraction of X-ray spectra.
The inner boundary condition of temperature, wind velocity and plasma $\beta$ are extrapolated from small-sacle GRMHD simulation of hot accretion flows focusing on the winds \citep{2021ApJ...914..131Y}. The radial grids of simulations spread a large dynamical range up to $10^6\rm~r_g$ and the latitudinal grids extend from 10$^{\circ}$ to 50$^{\circ}$ to avoid influence from jet and inflow. Only the toroidal magnetic field is taken into account since poloidal field can be negligible at large radii. The emissivity-weighted radial profiles of density and temperature of the simulated wind for the two sources can be fitted by:
\begin{equation}
\label{eq:denWm81}
n_{\rm wind,M81}\simeq1.7\times10^5\left(\frac{r}{3000\rm~r_{g}}\right)^{-2}\rm~cm^{-3}
\end{equation}
\begin{equation}
\label{eq:denWngc7213}
n_{\rm wind,NGC\,7213}\simeq2.7\times10^6\left(\frac{r}{2000\rm~r_{g}}\right)^{-2}\rm~cm^{-3}
\end{equation}

\begin{equation}
\label{eq:tWm81}
T_{\rm wind, M81}\simeq34.7\left(\frac{r}{3000\rm~r_{g}}\right)^{-1.28}\rm~keV 
\end{equation}
\begin{equation}
\label{eq:tWngc7213}
T_{\rm wind, NGC\,7213}\simeq127.5\left(\frac{r}{2000\rm~r_{g}}\right)^{-1.29}\rm~keV 
\end{equation}

Combining the derived average wind temperature with the theoretically predicted temperature profile above (Eq.\ref{eq:tWm81} - \ref{eq:tWngc7213}), we can determine the location of wind producing certain emission lines. The best fit temperature of the two hot wind components in NGC\,7213 are $16^{+6}_{-6}$ keV from HETG and $0.20^{+0.07}_{-0.05}$ keV from RGS, corresponding to $\sim8\times10^3~r_g$ ($\sim$0.04 pc) and $\sim3\times10^5~r_g$ ($\sim$1.4 pc) away from central black hole respectively. For M81*, the wind components detected in HETG hard X-ray bands locates around $\sim1\times10^4\rm~r_g$ ($\sim$0.04 pc).
The RGS spectrum of M81* shows a prominent blue-shifted Ne X Ly$\alpha$ line, while the blue-shifted O VIII Ly$\alpha$ line is marginal.
The line luminosity ratio between prominent blue-shifted Ne X Ly$\alpha$ ($5.0_{-2.3}^{+2.5}\times10^{37}$ erg/s) and postulated blue-shifted O VIII Ly$\alpha$ (\textless $5.4\times10^{37}$ erg/s with 3$\sigma$ upper limit) is found to be smaller than that collisional ionization equilibrium plasma could ever reach.
By adopting a sub-solar abundance of O and super-solar abundance of Ne, the temperature of the outflowing component can be constrained to $0.44^{+0.13}_{-0.09}$ keV, corresponding to the wind location of $9\times10^4\rm~r_g$ (0.3 pc).

In the soft X-ray regime, self-absorption of hot wind could not be neglected.
Given the gas temperature, density and radial velocity of hot wind from simulations, following the method described in Appendix \ref{appendix:opdepth}, we illustrate the total optical depth of hot wind self-absorption (both bound-bound and bound-free) over 0.6 - 1.1 keV band in Figure \ref{fig:opdepth}.
For M81*, the optical depth peaks at $\sim0.79$ for O VIII line and $\sim0.16$ for Ne X line, assuming the gas turbulent velocity is 200 km/s. 
For NGC\,7213, the peak optical depth of H-like O and Ne lines are $\sim0.26$ and $\sim0.04$.
Hot wind itself would be almost optically thick for O VIII Ly$\alpha$ line in M81*. 
After calibrating for self-absorption, predicted 3-$\sigma$ upper limit of ratio between O VIII and Ne X Ly$\alpha$ is $\gtrsim2.1$, corresponding to wind temperature $\gtrsim0.6$ keV, also located within $9\times10^4\rm~r_g$ (0.3 pc), consistent with the location determined above.

\begin{figure}
\centering
\includegraphics[width=0.8\linewidth]{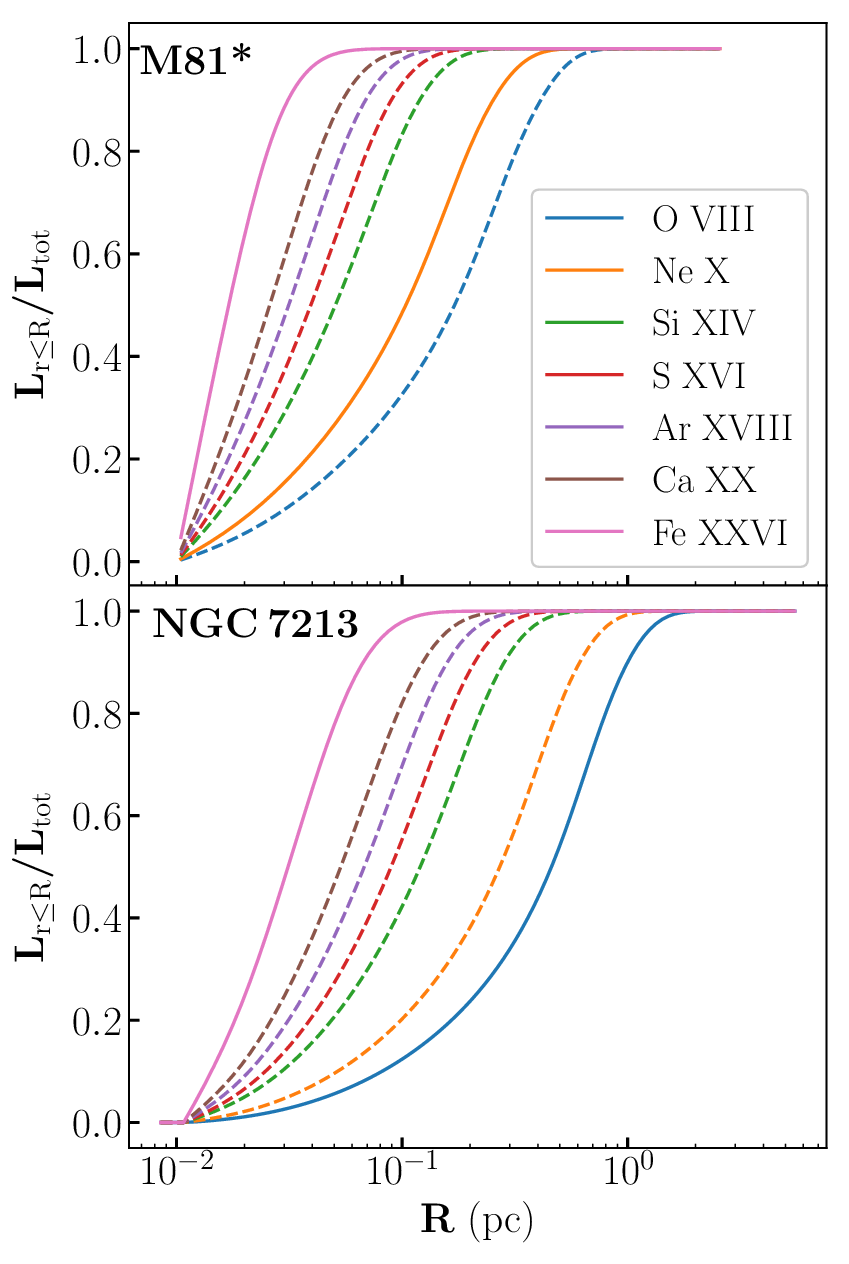}
\caption{
Ratio between the accumulative luminosity within radius $R$ and total luminosity of emission lines produced by hot winds as a function of the distance $R$ to central black hole for M81* (upper panel) and NGC\,7213 (lower panel), predicted by numerical simulations.
Highly ionized Fe and O, Ne lines with clear detection in observations are marked in solid lines. 
Emission lines with elusive detection are marked in dashed lines.
\label{fig:growthC}
}
\end{figure}

Luminosity of H-like Ne produced by putative free-expanding hot wind is predicted to be $\sim1.3\times10^{37}$ erg/s in M81* and $\sim5\times10^{39}$ erg/s for O VIII Ly$\alpha$ in NGC\,7213.
Compared with the observed luminosity ($5^{+3}_{-2}\times10^{37}$ erg/s for M81* and $4\pm3\times10^{38}$ erg/s for NGC\,7213, extra enhancement is required in M81*.
As the temperature of the outflowing hot wind continuously decreases, we would expect to detect blue-shifted characteristic emission lines from highly ionized Si, S, Ar, Ca , etc., produced by wind with intermediate temperature between Fe-emitting and Ne/O-emitting regions.
Only upper limits of these lines can be estimated from observed spectra, although they are still consistent with the prediction from simulations shown in Table \ref{tab:lineLwind}.
Figure \ref{fig:growthC} shows the growth curve of different emission lines predicted by the hot wind simulations.

In the scenario of free-expanding wind, wind velocity is expected to remain almost constant or slightly increase with distance \citep{2020ApJ...890...80C,2020ApJ...890...81C}.
However, there are discrepancies in gas velocity between the outflowing plasma detected in hard and soft X-ray bands, as shown in Table \ref{tab:hwProp}. Specifically, for M81* gas velocity drops from $2.8\pm0.2\times10^3$ km/s to $1.9^{+0.6}_{-0.6}\times10^3$ km/s, while for NGC\,7213 it increases from $1.2^{+0.1}_{-0.2}\times10^3$ km/s to $2.6_{-0.4}^{+0.7}\times10^3$ km/s.
Moreover, the simulated spectral model for free-expanding hot wind derived in \citetalias{2021NatAs...5..928S} and \citetalias{2022ApJ...926..209S}, which well fit the HEG spectra over 5-8 keV band, fail to adequately describe the observed spectra in soft X-ray range. This could be because the number density of the observed outflowing components do not follow the radial profile predicted by adiabatic expansion wind, as shown in Appendix \ref{appendix:hwsimu} and Figure \ref{fig:hwModSimu}.
This indicates that free expansion of hot winds alone cannot simultaneously explain the emission lines detected in hard and soft X-ray bands. 

\begin{deluxetable}{c|lccc}
\tablecaption{Properties of Wind Components\label{tab:hwProp}}
\tablehead{
\colhead{Source} & \colhead{$v_{\rm wind}$} &
\colhead{$r_{\rm aver}$} & \colhead{$t_{\rm aver}$} & \colhead{$<n_{\rm aver}>$}\\
\colhead{(1)} & \colhead{(2)} & \colhead{(3)} & \colhead{(4)} & \colhead{(5)}}
\startdata
\multirow{2}{*}{M81} & $2.8\pm0.2$ & $\sim0.04$ & $\sim10$ &  $4.8\times10^4$\\
& $1.9\pm0.6$ & $\lesssim0.3$ & $\sim98$ & $5.2\times10^2$\\
\hline
\multirow{2}{*}{NGC\,7213} & $1.2^{+0.1}_{-0.2}$ & $\sim0.04$ & $\sim6$ & $1.1\times10^{5}$\\
& $2.6_{-0.4}^{+0.7}$ & $\sim1.4$ & $\sim196$ & $7.0\times10^1$\\
\enddata
\tablecomments{
(1) Source name. For each source, parameters given in the top (bottom) row are deduced from HETG(RGS) observations.
(2) Velocity of the wind in units of $\rm 10^3~km/s$.
(3) Wind location determined by comparing observed average plasma temperature with (extrapolated) temperature radial profile from hot wind simulations, in units of pc.
(4) Dynamical timescales of winds reaching $r_{\rm aver}$ in units of yr.
(5) Estimated number density of the wind at $r_{\rm aver}$ in units of $\rm cm^{-3}$ derived from line diagnostics and spectral modeling for M81 and NGC\,7213, assuming the wind opening angle is $\sim45^{\circ}$ and density decays with $r^{-2}$.
}
\vspace{-3em}
\end{deluxetable}

\subsection{Shocked circumnuclear medium}\label{subsec:shock}
A more advanced consideration is that, as hot wind propagates outward it may encounter circumnuclear medium before its temperature dropping to the point where emission lines of Si XIV, S XV, etc. can be produced ($\sim0.1$ pc from the growth curve Figure \ref{fig:growthC}).
When the supersonic hot wind interacts with the circumnuclear medium, it will be shocked, compressed and heated to $T\gtrsim6.5\times10^{7}(\frac{v_{\rm shock}}{\rm 2000~km/s})\rm~K$, assuming a strong shock limit.
Radiative cooling should be efficient in the shocked medium since it has been compressed to higher density. Its temperature will then decrease to an appropriate range ($10^{6-7}$ K) to generate highly ionized O VIII and Ne X lines.
Velocity of such shocked medium would be different from the freely propagating hot wind.
This scenario could explain the absence of emission lines between highly ionized Fe and O, Ne as well as the observed discrepancies in properties between the hotter and cooler blueshifted gas components. 

\begin{figure*}
\centering
\includegraphics[width=0.45\linewidth]{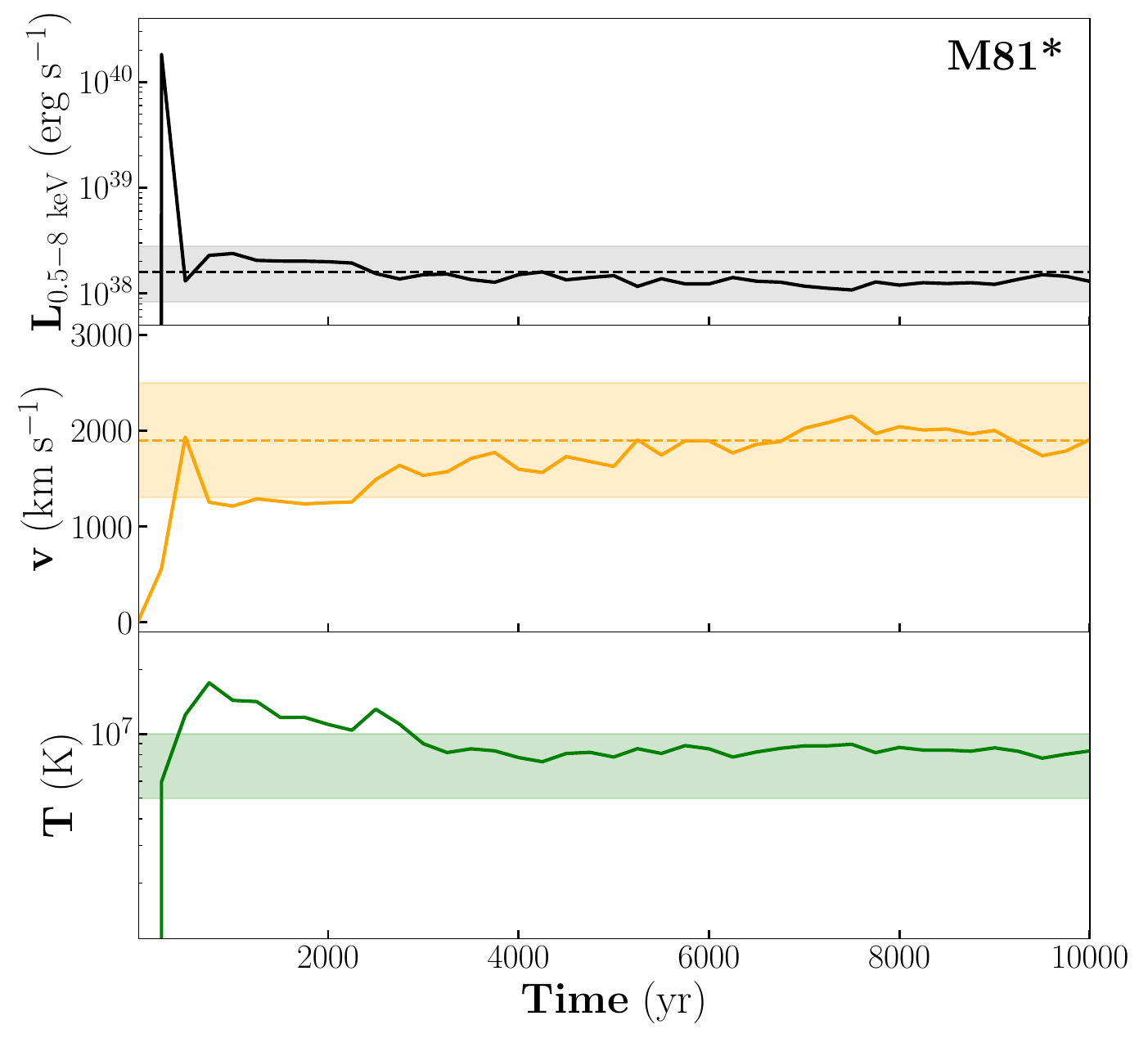}
\includegraphics[width=0.44\linewidth]{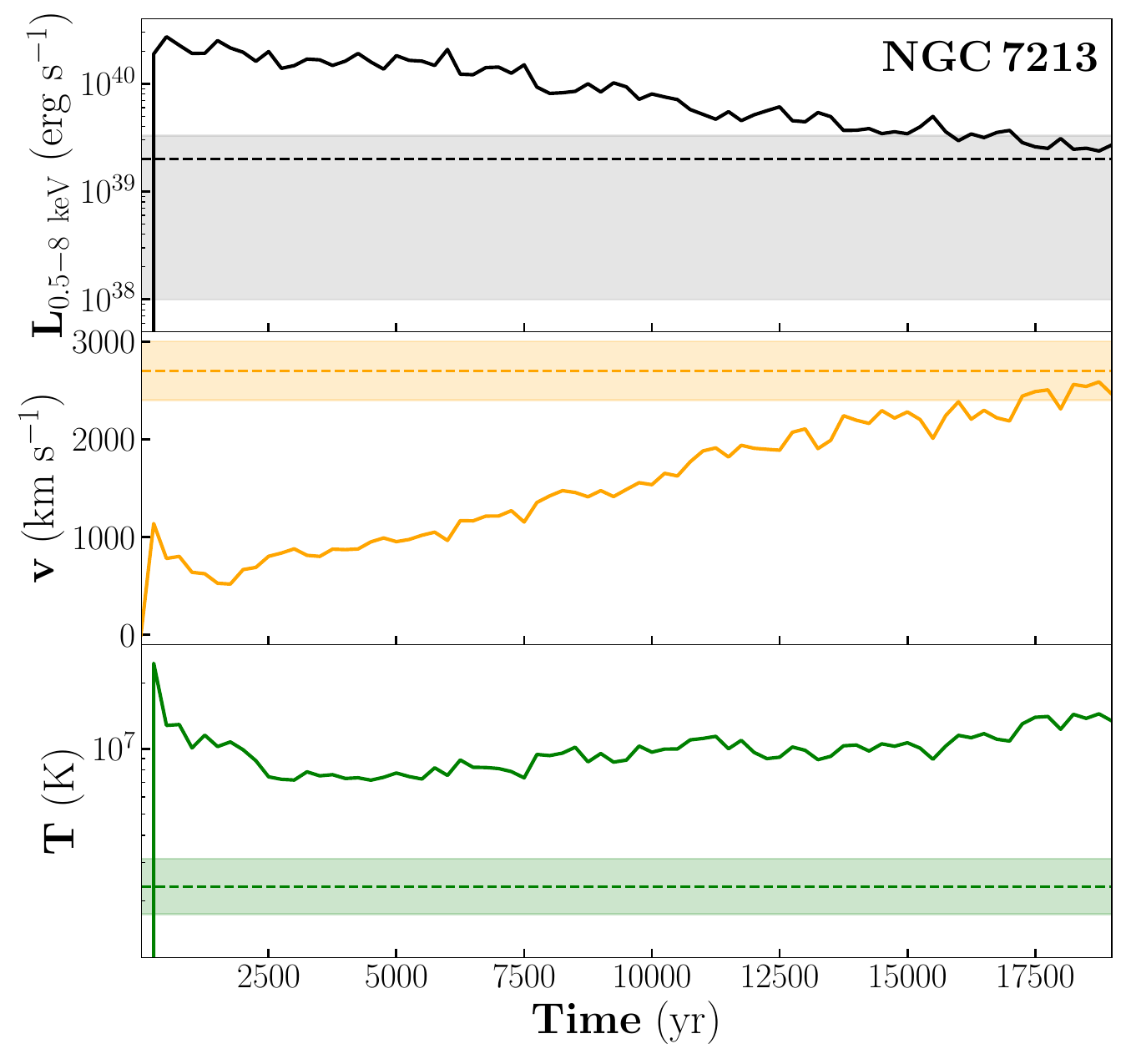}
\caption{Predicted 0.5-8 keV X-ray luminosity (top panel), emissivity-weighted average velocity (middle panel) and temperature (bottom panel) of circumnuclear medium shocked by the expanding hot winds within central 10 pc for M81* and NGC\,7213. Dashed lines and shadow regions in the top and middle panel mark the observed value and 90\% error bar for the blue-shifted gas component. Shadow region in the bottom panel give the temperature constraints from line ratios. Continuously injection of hot winds cause the shocked medium retains hot ($\sim10^6$-$\sim10^7$ K) and keeps going outward with speed about $\sim10^3$ km/s.
\label{fig:shock}
}
\end{figure*}

\begin{figure*}
\centering
\includegraphics[width=0.48\linewidth]{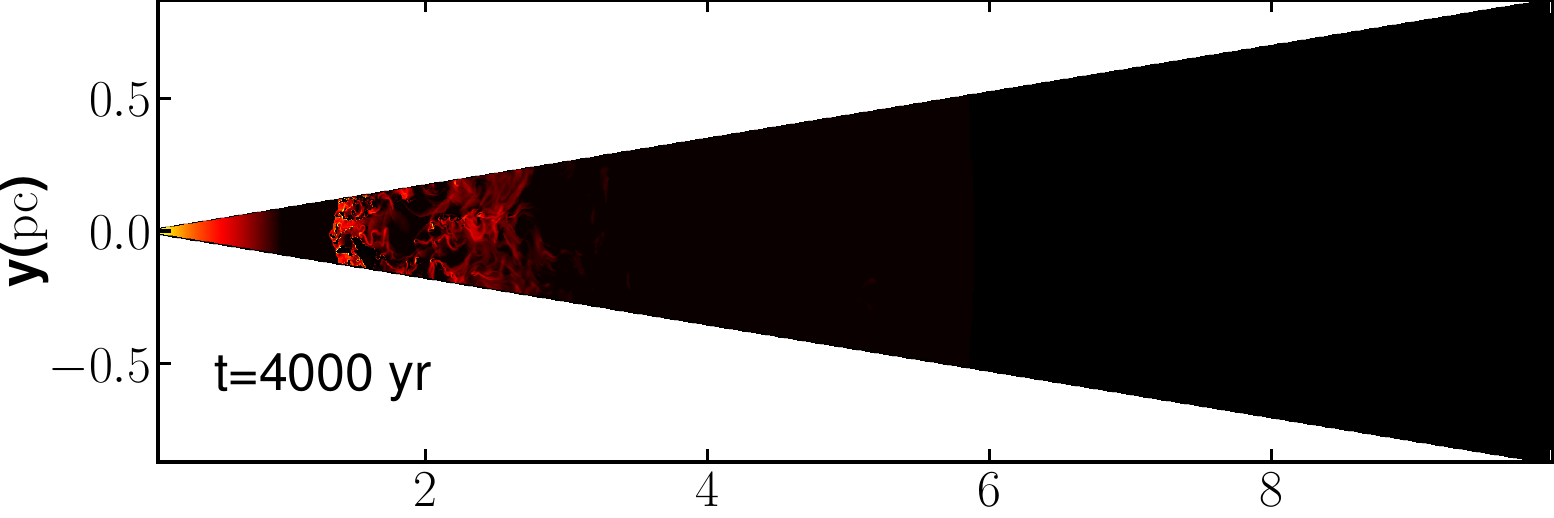}\includegraphics[width=0.47\linewidth]{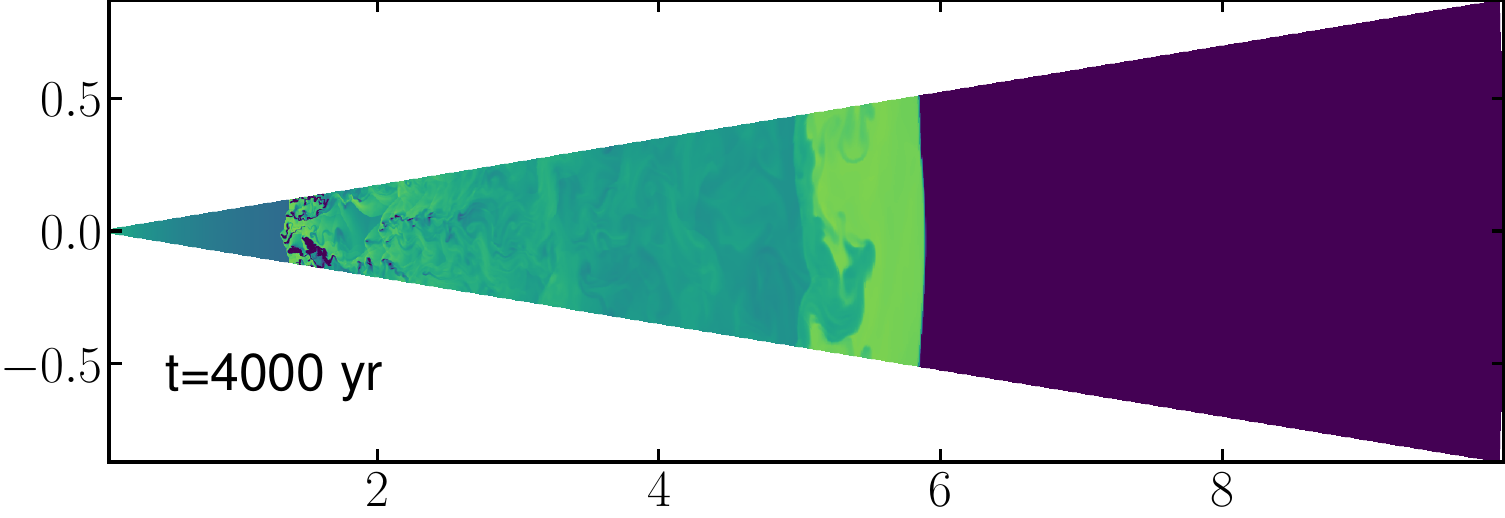}
\includegraphics[width=0.48\linewidth]{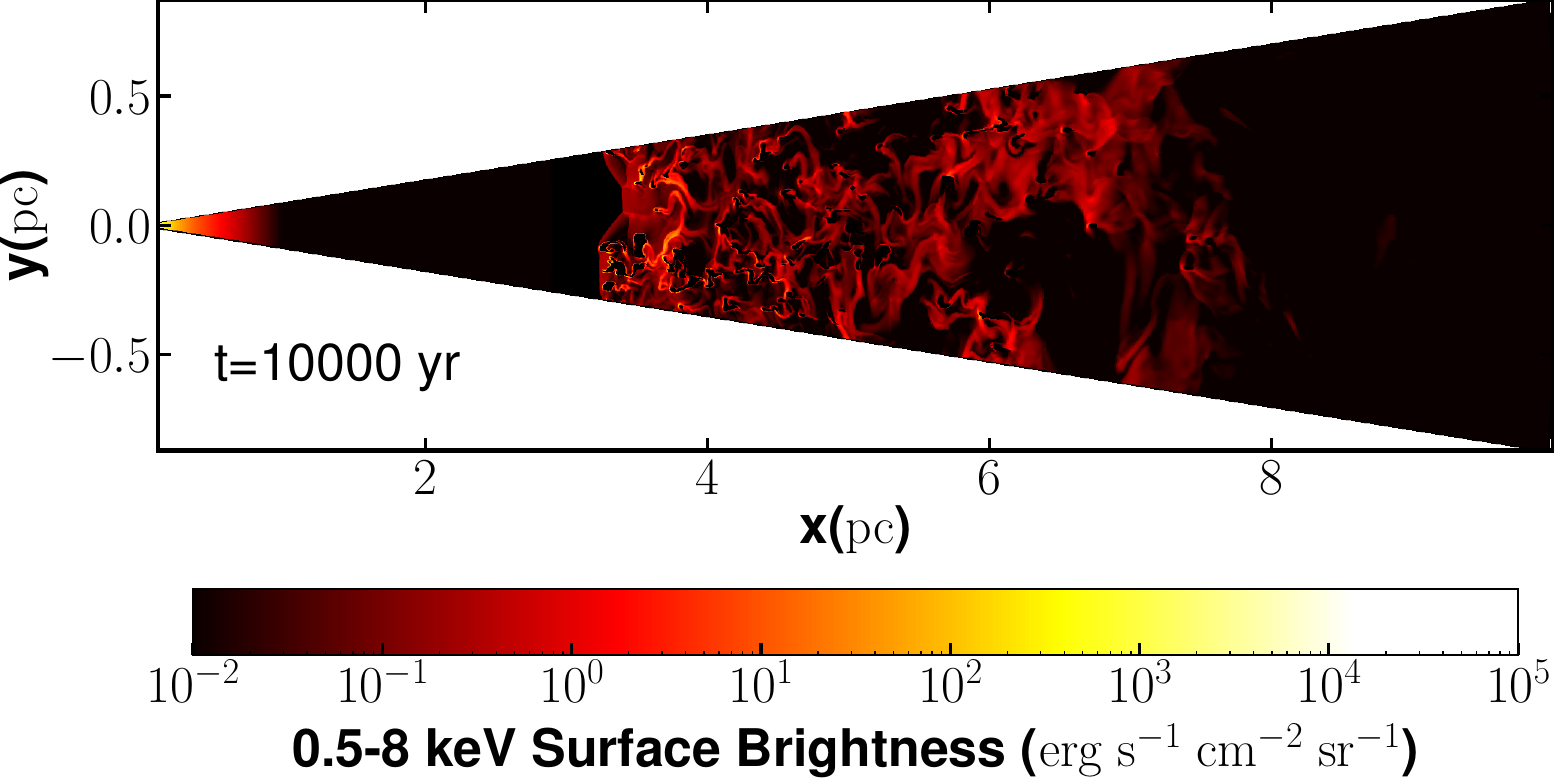}\includegraphics[width=0.47\linewidth]{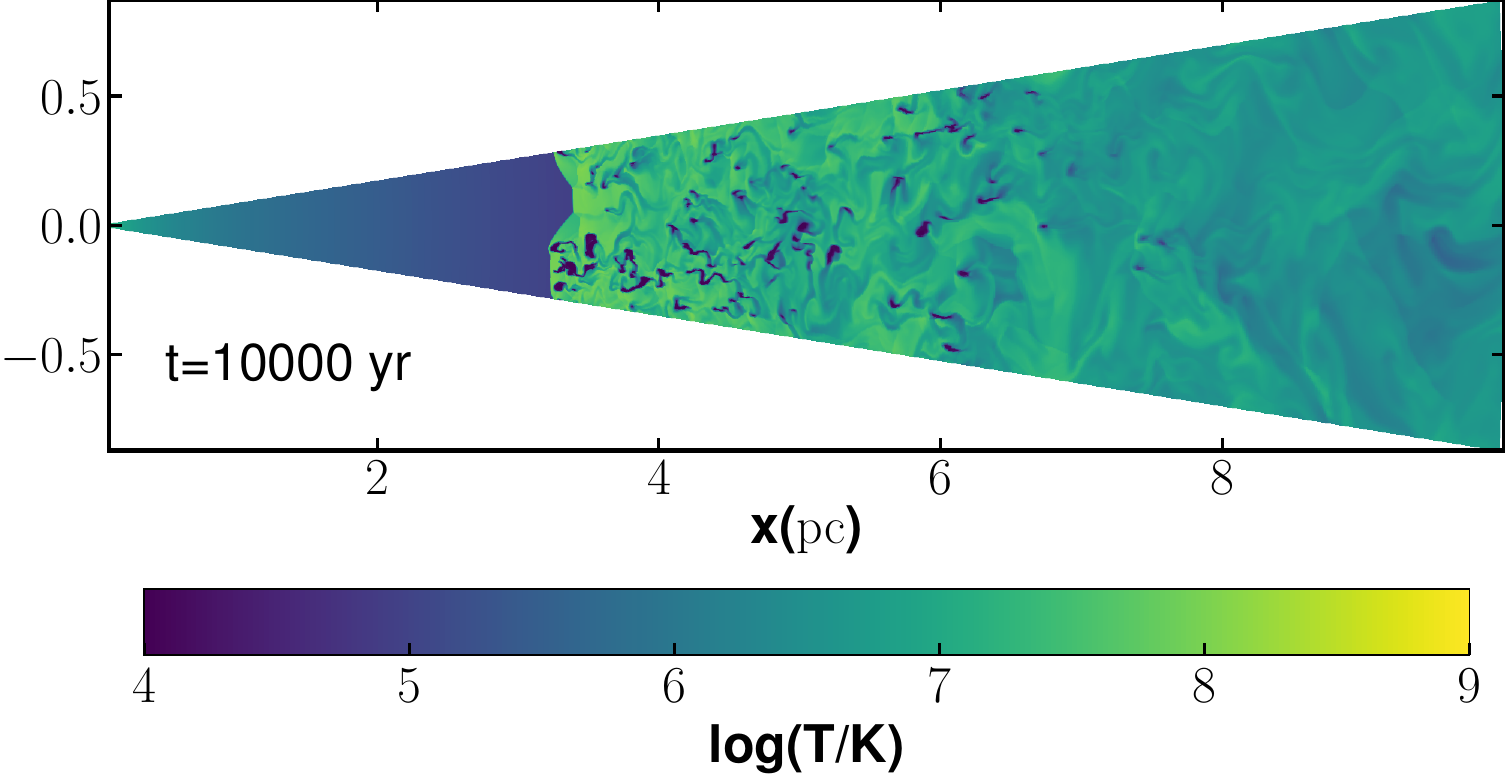}
\caption{Predicted 0.5-8 keV X-ray {\bf surface brightness (left columns)} and temperature distribution (right columns) of circumnuclear medium shocked by the expanding hot winds within central 10 pc for M81* at {\bf t=4000 yr (upper panels)} and t=10000 yr (lower panels). Strong X-ray emitting regions coincide with $\sim10^7$ K gas. Hot winds continuously push the circumnuclear medium outward.
\label{fig:shock_X-ray}
}
\end{figure*}

We perform hydrodynamical simulations to examine this scenario for M81* and NGC\,7213.
First, we simply consider setting the circumnuclear medium in homogeneously and spherically distribution around the central black hole. Using publicly available code {\sc pluto} \citep[version 4.4;][]{2007ApJS..170..228M}, we adopt a two-dimensional, static spherical grid. The grid is logarithmically distributed in $R$ direction and uniformly distributed in $\theta$ direction with $4096\times312$ cells in the domain $\rm [0.1pc,10pc]\times[85^\circ,95^\circ]$.
Hot winds are injected from the inner boundary, temperature of which are calculated from Eq. \ref{eq:tWm81} - \ref{eq:tWngc7213}.
Density of the hot winds in NGC\,7213 also follows the derived profile in Eq. \ref{eq:denWngc7213} but is increased to 4 times of that in Eq. \ref{eq:denWm81} for M81*.
The intrinsic emissivity-weighted average radial velocity of hot winds from \citetalias{2021NatAs...5..928S} and \citetalias{2022ApJ...926..209S} is adopted, which is $4000\rm~km~s^{-1}$ for M81* and $7000\rm~km~s^{-1}$ for NGC\,7213.
Outflow condition is set for the outer radial boundary and periodic boundary condition is adopted in the $\theta$-direction.
We include radiative cooling using cooling function for solar abundance plasma, according to the lookup table presented by \citet{2020MNRAS.497.4857P} based on photoionization code {\sc Cloudy} \citep{2017RMxAA..53..385F}.
Gravity and magnetic field are ignored for simplicity.

The $T\sim5000~$K circumnuclear medium is assumed to be uniformly distributed with density $n_{\rm H}=2\times10^4\rm~cm^{-3}$ for M81* and $n_{\rm H}=2\times10^5\rm~cm^{-3}$ for NGC\,7213 whose accretion rate is 10 times higher.
The injected hot winds begin to encounter with the medium at 0.1 pc away from black hole and are allowed to evolve for $\sim10^4\rm~yr$, long enough for the shocked medium to cool through radiation.
Time evolution of the simulated X-ray emissivity-weighted average radial velocity and temperature of the shocked medium are shown in Figure \ref{fig:shock}.
The total X-ray luminosity of the shocked medium has been calculated and scaled from the 2D simulation to 3-dimension scenario assuming axis symmetry with $2\pi$ solid angle for the blue-shifted half sphere alone. 
The initial setups of the simulations are summed up in Table \ref{tab:initsimu}.

For M81*, by setting $2\times10^3\rm~M_{\rm\odot}$ of warm ionized medium within central 1 pc,
such post-shocked medium could cool down to an average temperature $\sim7\times10^6$ K and reaches average velocity of $\sim2000$ km/s. The estimated 0.5-8 keV X-ray luminosity of the shocked medium is $\sim1.5\times10^{38}$ erg/s, scaled from the 10-deg slice to spherical distribution.
This is consistent with the blue-shifted component derived from observed spectra ($1.6_{-1.1}^{+8}\times10^{38}$ erg/s), as shown in Figure \ref{fig:shock}.
And strong X-ray emitting regions approximately coincide with $\sim10^7$ K medium, as shown in Figure \ref{fig:shock_X-ray}.
Assuming the Ne is twice as solar abundance, the estimated blue-shifted Ne X Ly$\alpha$ from shocked medium is $6\times10^{37}$ erg/s, consistent with the observed value.
The cooling timescale of the shocked medium is $\sim3000$ yrs and is smaller than the dynamical timescale ($\sim 5000$ yr) for hot wind reaching 10 pc.

Similarly for NGC\,7213,
by setting $\sim2\times10^4\rm~M_{\odot}$ of warm ionized gas placed within 1 pc, the shocked medium could reach $\sim3000$ km/s with 0.5-8 keV luminosity of $10^{39}\rm~erg~s^{-1}$ within $2\times10^4$ yrs (Figure \ref{fig:shock}). 
Simulated temperature of the shocked medium seems several times higher than the best-fit blueshifted gas temperature in NGC\,7213. The observed temperature is constrained mainly through the detection of single blueshifted O VIII line and upper limits of Ne and other highly ionized lines, thus may invoke large uncertainties.
Overall, the blue-shifted emission lines detected in soft X-ray bands could be the footprints of interaction between hot winds and circumnuclear warm ionized gas.

An alternative and more realistic approach to modeling the surrounding medium within the pc-scale is to consider a clumpy morphology instead of homogeneous and continuous distribution. Clumpy clouds can achieve a larger covering factor with less amount of gas.
In this case, we replace the continuously distributed gas with various numbers of randomly distributed clumpy clouds, each with a radius of 0.1 pc and a temperature of $\sim$1000 K. This simulation has been performed in a 2-dimensional box with Cartesian coordinate using {\sc FLASH} v4.6 code \citep{2000ApJS..131..273F}.
Hot winds are injected at the inner spherical boundary around black hole at the box center.
Temperature and density of the injected wind exactly follow the results from previous simulations in Eq. (\ref{eq:denWm81}) - (\ref{eq:tWngc7213}).
Detailed simulation setups are described in Appendix \ref{appendix:clump}.
At $\sim10^3$ yr and after the hot winds passing through all the clumps, emission from shocked clumpy clouds dominates the 0.5-8 keV X-ray luminosity.
The simulation results are consistent with the observed gas temperature and velocity as shown in the T-$v_{\rm r}$ phase plot in Fig. \ref{fig:flash_m81} and Fig. \ref{fig:flash_n7213} as well.
By introducing $90\rm~M_{\odot}$ and $1350\rm~M_{\odot}$ of warm ionized gas within pc-scale for M81* and NGC\,7213, 
the shocked clumpy clouds can achieve comparable X-ray luminosity with observations in soft X-ray bands.

Multi-wavelength observations indicate the existence of sufficient warm ionized circumnuclear medium within pc-scale for M81* and NGC\,7213, lending support to the above hot wind interaction scenario.
For M81*, \citet{2019MNRAS.488.1199D} explains the complex kinematic structure of broad H$\alpha$ emission line profile with a dynamically unstable, infalling photoionized H$\rm^+$ gas with $10^4$-$10^5\rm~cm^{-3}$ density and large covering factor that is bounded to the inner 0.9 pc.
Mass of this pc-scale warm ionized gas is estimated to be $\sim500\rm~M_{\odot}$.
Characteristic forbidden and semi-forbidden emission lines presented in the UV spectrum imply the existence of shock, either by jet proposed by \citet{2019MNRAS.488.1199D}, or by hot winds as we suggest.
A double-peaked broad H$\alpha$ component has been spotted in the optical spectra of NGC\,7213 nucleus (\textless 110 pc) taken from Gemini Multi-Object Spectrograph.
Flux of this broad component varies up to 30\% within 4 months and up to 10\% within 28 days, probably indicating that the gas component is located within $\sim1\rm~pc$ from the central engine \citep{2014MNRAS.438.3322S,2017MNRAS.472.2170S}. From the average integrated luminosity of this broad H$\alpha$ component ($\sim3\times10^{40}\rm~erg~s^{-1}$), considering $M_{\rm ion}=\mu m_{\rm p}\frac{L_{\rm H\alpha}}{n_{\rm e}j_{\rm H\alpha}}$, the mass of the circumnuclear warm ionized gas is estimated to be $\sim5\times10^3\rm~M_{\odot}$ assuming the average density within 10 pc is $2\times10^4\rm~cm^{-3}$,
roughly consistent with that required in our shocked medium scenario. Here
$j_{\rm H\alpha}=3.56\times10^{-25}\rm~erg~cm^3~s^{-1}$ is the effective emissivity for case B recombination \citep{2006agna.book.....O}. In the simulations, we initially assign the amount of circumnuclear medium and the outer boundary has been set to outflowing condition.
While in real case,
the origin of the circumnuclear medium could be external, like infalling gas clouds that cool down {\it in-situ} due to thermal instability\citep{2012MNRAS.424..728B} or stellar wind of nuclear star clusters.
These surrounding medium may chaotically but continuously keep falling in and interact with hot wind at pc scale.

The continuously injected hot winds provide both energy and momentum feedback on the circumnuclear gas. In the energy feedback aspect, hot winds act as a heating source by shock heating the  gas in the central pc-scale. Shocked medium can maintain temperature of $\gtrsim10^7$ K. In the momentum feedback aspect, the wind can kinetically impede the infalling of interstellar medium, as suggested in Fig.\ref{fig:shock} and Fig.\ref{fig:shock_X-ray}.
To quantitatively examine the importance of momentum feedback by hot wind, we estimate the ram pressure force $\rho_{\rm wind} v_{\rm wind}^{2}S$ that hot winds would exert on a cloud and compare it with the gravitational force of the cloud from the black hole. The radius of cloud is assumed to be 0.1 pc and its distance to the black hole ($\sim10^8\rm~M_{\odot}$) is assumed to be 1 pc.
Wind density $\rho_{\rm wind}\simeq n_{\rm wind}\mu m_{p}$ at the cloud front is determined from density profile of free-expanding wind for simplification.
Number density of the cloud is chosen to be $\sim10^4\rm~cm^{-2}$, consistent with the H$\alpha$ observation.
The ram pressure force that the hot winds provide on each single cloudlet is found to be $\sim9\times10^{25}$ N, large enough to compete with the gravitational force, which is $\sim2\times10^{25}$ N. The ratio of the ram pressure force the hot wind exert on the hot phase of the circumnuclear gas and the gravitational force would be larger. This estimation suggests that hot wind would be able to hinder the gas accretion onto the central black hole and keep the central AGN dim. Such a role of hot wind in AGN feedback is fully consistent with the detailed numerical simulation result obtained by \citet{2019ApJ...885...16Y}.
 
\section{Summary}
\label{sec:sum}

In this paper, we have analyzed XMM-Newton RGS high-resolution spectra of two LLAGNs with previously detected  $T\gtrsim10$-keV hot wind launched by hot accretion flows, M81* and NGC\,7213. The aim of the present paper is to focuse on the propagation of hot winds and their interaction between hot winds and circumnuclear ISM. Our main results are as follows:
\begin{itemize}
\item[1.]{We detect blueshifted Ne X Ly$\alpha$ (O VIII Ly$\alpha$) emission line in M81* (NGC\,7213) from the 1-order RGS spectra in 0.5-2 keV band, which correspond to outflowing plasma components with temperature of 0.44 keV (0.2 keV) and velocity $1.9\pm0.6\times10^3$ ($2.6^{+0.7}_{-0.4}\times10^3$) km/s.}
\item[2.]{
Given that blue-shifted emission lines produced by temperature larger than 10 keV have been detected and identified as wind launched from hot accretion flow in these two sources, we first associate the newly detected emission lines in this paper with hot wind freely propagating to larger radii. However, we find this scenario fails to explain the detected velocity of the outflow and the intensity of the emission lines. It is also hard to explain the absence of some characteristic emission lines for intermediate temperature between Fe and O, Ne. 
}
\item[3.]{
Considering the presence of circumnuclear gas around the AGN, we have performed numerical simulations by considering the interaction between the hot wind and the clumpy circumnuclear gas. We find that the temperature, velocity, and density of  circumnuclear gas shock-heated by the hot wind around pc-scale can well explain the newly observed blue-shifted emission lines. }
\item[4.] {The above result provides strong evidence for hot wind heating the circumnuclear gas. In addition to this energy feedback,
we have also evaluated the momentum feedback effect by estimating the ram pressure force the hot wind exerts on the circumnuclear clumpy clouds at pc-scale. It is
comparable or larger than the gravitational force due to central black hole, thus the wind can impede the accretion of gas. The energy together with the momentum feedback imposed by the hot wind keeps the AGN in a low-luminosity state. This result is consistent with numerical simulations \citep{2019ApJ...885...16Y}.}
\end{itemize}

\begin{acknowledgments}
\twocolumngrid
We would like to thank Luis C. Ho, Xinwen Shu, Junjie Mao and Bocheng Zhu for helpful discussions. FS is supported in part by the China Postdoctoral Science Foundation (grants 2022TQ0354 and 2022M723279). FY is supported by Natural Science Foundation of China (grants 12133008, 12192220, 12192223) and China Manned Space Project (CMS-CSST-2021-B02).
The simulations have made use of the High Performance Computing Resource at Shanghai Astronomical Observatory and Nanjing University. This research has made use of data obtained from the Chandra Data Archive and the XMM-Newton Science Archive, and software provided by the Chandra X-ray Center (CXC) in the application package CIAO and the ESAC Science Data Centre with package SAS.
\end{acknowledgments}

%

\vspace{5mm}
\facilities{XMM-Newton (RGS), CXO}


\software{astropy \citep{2013A&A...558A..33A,2018AJ....156..123A}, 
          Xspec \citep{1996ASPC..101...17A},
          CIAO \citep{2006SPIE.6270E..1VF},
          SAS \citep{2004ASPC..314..759G},
          PLUTO \citep{2007ApJS..170..228M},
          FLASH \citep{2000ApJS..131..273F}
          }



\appendix
\twocolumngrid
\renewcommand{\thefigure}{\Alph{section}\arabic{figure}}
\renewcommand{\thetable}{\Alph{section}\arabic{table}}
\section{Off-nucleus contamination}\label{appendix:contam}
Unlike transmission gratings, RGS cannot disentangle the spatial information in the dispersion direction. Both off-nucleus sources like X-ray binaries or distant AGNs
and spatial extension of unresolved diffuse emission 
could potentially bring apparent blue-shifted emission lines to the extracted spectrum. 
Spatial offset $\Delta\theta$ (in unit of arcmin) in the dispersion direction would result in $\simeq\frac{0.138\Delta\theta}{m}$ (in unit of $\rm \AA$, m is the order of spectrum) offset in wavelength for certain emission line \footnote{\url{https://xmm-tools.cosmos.esa.int/external/xmm_user_support/documentation/uhb/rgsspecres.html}}.

Off-nucleus point sources can be clearly seen in the stacked 0-order image of Chandra HETG observations for the 2 sources.
The wavelength offset of detected blue-shifted emission lines corresponds to potential contamination sources within $\sim1$\arcmin (1 kpc for M81* and 6.6 kpc for NGC\,7213 in projection).
Spectra of point sources within this region around central AGN are identified, extracted and coadded together. 
In M81*, luminosity of Ne X Ly$\alpha$ line from the off-nucleus sources is estimated to be $5\pm4\times10^{36}$ erg/s, about an order of magnitude smaller than the observed blue-shifted Ne Ly$\alpha$ emission line $5^{+3}_{-2}\times10^{37}$ erg/s.
In NGC\,7213, only a 3-$\sigma$ upper limit of O VIII Ly$\alpha$ luminosity $\lesssim1.4\times10^{38}$ erg/s is obtained from contamination spectrum, which is still several times smaller than the observed blue-shifted O VIII luminosity $4\pm3\times10^{38}$ erg/s.
Therefore, the off-nucleus point sources could not account for the detected blue-shifted lines in RGS spectra.

Standard RGS response matrix is built for point sources.
We convolve our models with {\it rgsxsrc} script embedded in {\it Xspec} to test the spatial broadening effect of diffuse emission on the spectral modeling. 
Images of 0.5-2 keV band with off-nucleus point sources masked are generated from Chandra observations as the required input angular structure function.
We do not find significant changes between best-fit values with those in Table \ref{tab:continuum} and Table \ref{tab:emisLine}.

\setcounter{figure}{0}
\begin{figure*}
\centering
\includegraphics[width=0.75\linewidth]{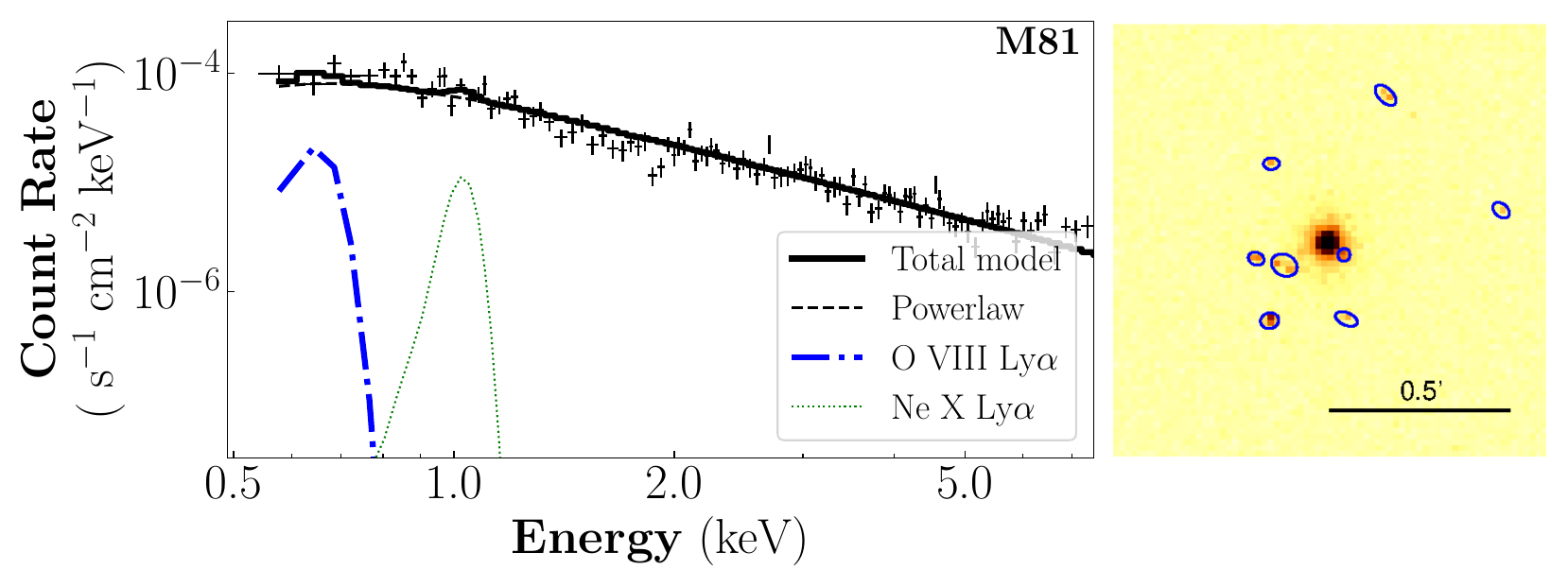}
\includegraphics[width=0.75\linewidth]{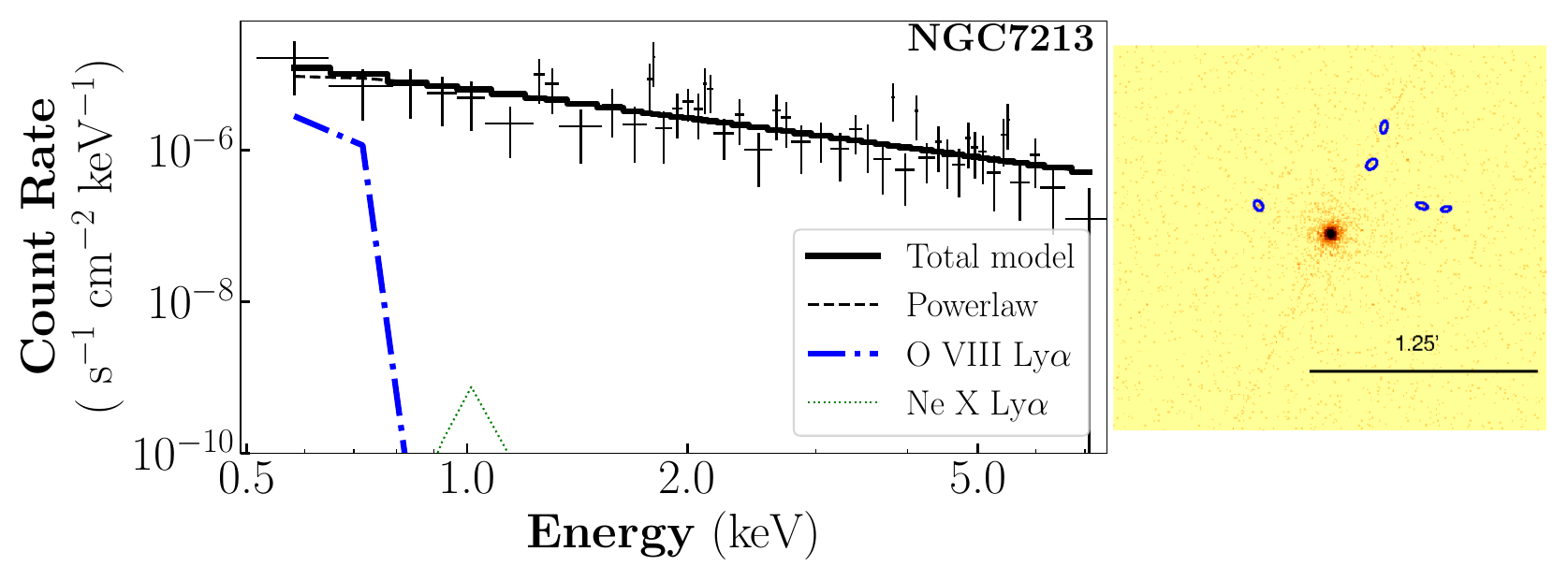}
\caption{
Stacked Chandra HETG 0-order image and spectra of off-nucleus contamination point sources in M81* (top panel) and NGC\,7213 (bottom panel). 
Black crosses mark the observed data. Two gaussian models are placed at putative line centroids, providing 3-$\sigma$ upper limit constraints to O VIII Ly$\alpha$ (blue dashed-dot lines) and Ne X Ly$\alpha$ (green dotted lines) from contamination sources. Luminosity of these contaminated lines is insufficient to account for the blue-shifted lines detected in RGS spectra.
Solid and dashed black lines describe the total and best-fit power-law continuum models of the stacked off-nucleus spectra.
Detected off-nucleus point sources (blue solid circle and ellipses) are marked in the 0.5-8 keV X-ray image.
\label{fig:off_nu}
}
\end{figure*}

\section{Absorption of the wind} \label{appendix:opdepth}
Hot wind self-absorption should be consist of bound-free continuum absorption and absorption due to certain bound-bound transition.
Optical depth caused by bound-free transition for ion {\it I} of certain element {\it X} can be calculated by $\tau_{\rm I}(E)=f_{\rm X}f_{\rm I}(T)n_{\rm tot}\sigma_{\rm I}(E)(1-e^{-\frac{E}{kT}})\Delta r$, where $f_{\rm X}$ is the abundance of element X and $f_{\rm I}$ is the ion fraction under certain gas temperature T. Gas number density is represented by $n_{\rm H}$ and $\sigma_{\rm I}(\nu)$ is the photon energy E dependent photoelectric absorption cross section. $\Delta r$ is depth of the gas. 
We sum over all ion spices and over all grids of a specific sight line to estimate total bound-free optical depth. We find that total optical depth does not vary significantly with sight lines, so inclination angle of 15$^{\circ}$ is chosen.
Energy dependent optical depth caused by transition between level $l$ and $u$ of ion I are expressed by $\tau_{\rm lu}(E)=n_{\rm H}f_{\rm X}f_{\rm I}(T)\frac{\pi q_{\rm e}^2}{m_{\rm e}c}f_{\rm lu}\phi(E)$, where $q{\rm e}$ and $m_{\rm e}$ are charge and mass of electron. $f_{\rm lu}$ is the oscillator strength of the transition.
The line profile described by Voigt function $\phi(E)$ is a function of Doppler broadening velocity, energy deviation from line centroid and natural broadening damping factor $\gamma$. Doppler broadening velocity $b_{\rm v}$ is a combination of thermal broadening and turbulent motion $(\frac{2kT}{m_{\rm i}}+v_{\rm turb}^2)^{1/2}$\citep{1979rpa..book.....R}.
Atomic data like photoelectric absorption cross section and temperature dependent ion fraction are taken from ATOMDB\textsuperscript{\ref{atomdb}}.

\setcounter{figure}{0}
\begin{figure*}
\centering
\includegraphics[width=0.49\linewidth]{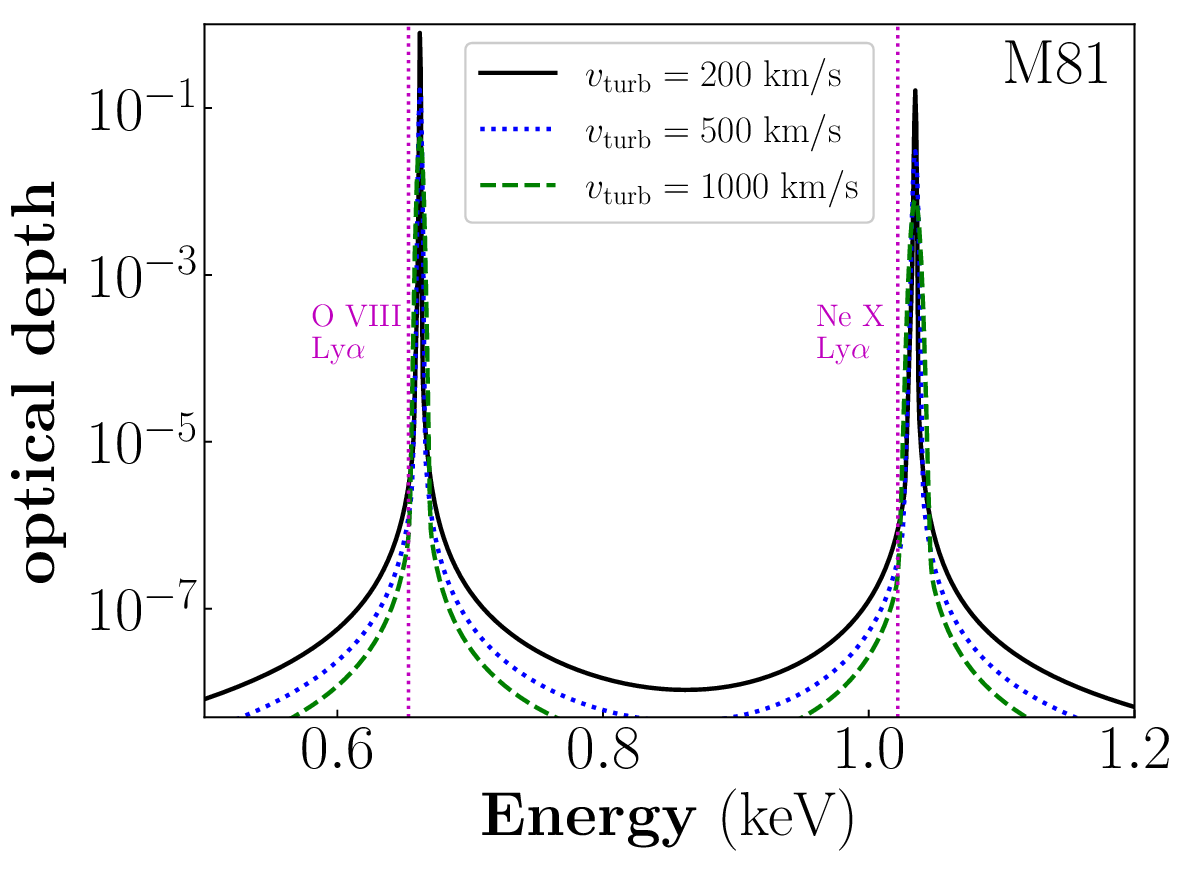}\includegraphics[width=0.49\linewidth]{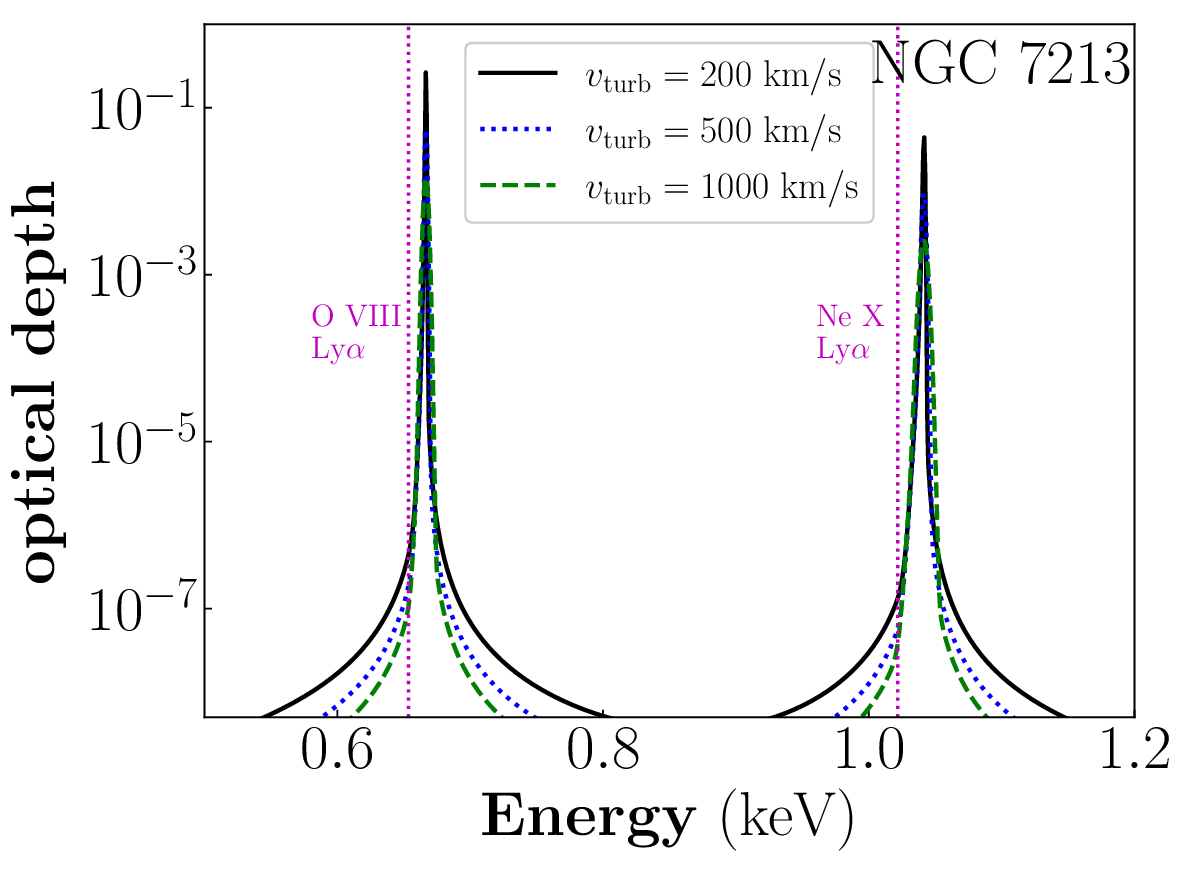}
\caption{
Optical depth of both bound-bound and bound-free self absorption from simulated hot winds for M81* (left panel) and NGC\,7213 (right panel) with inclination angle of 15$^\circ$, assuming different turbulent velocity. Vertical magenta dashed lines mark the rest-frame central energy of O VIII Ly$\alpha$ and Ne X Ly$\alpha$ emission lines.
Hot wind self-absorption may suppress the intensity of O VIII Ly$\alpha$ in M81*, while remains reasonably negligible for Ne X Ly$\alpha$ and in NGC\,7213.
\label{fig:opdepth}
}
\end{figure*}

\setcounter{table}{0}
\begin{deluxetable}{c|l|ccc}
\tablecaption{Predicted and observed blue-shifted characteristic line luminosity from hot winds\label{tab:lineLwind}}
\tablehead{
\colhead{Source} & \colhead{Line} & \colhead{$L_{\rm RGS}$} & \colhead{$L_{\rm HEG}$} &
\colhead{$L_{\rm simu}$} \\
\colhead{(1)} & \colhead{(2)} & \colhead{(3)} & \colhead{(4)} & \colhead{(5)}
}
\startdata
\multirow{6}{*}{M81*} & O VIII Ly$\alpha$ & \textless5.4 & - & 3.0 \\
& Ne X Ly$\alpha$ & $5_{-2}^{+3}$ & \textless3.0 & 1.0  \\
& Si XIV Ly$\alpha$ & -  & \textless1.5  & 1.1 \\
& S XVI Ly$\alpha$ & -  & \textless2.1  & 0.8 \\
& Ar XVIII Ly$\alpha$ & - & \textless1.7  & 0.3 \\
& Ca XX Ly$\alpha$ & - & \textless2.2  & 0.2 \\
\hline
\multirow{6}{*}{NGC\,7213} & O VIII Ly$\alpha$ & $4\pm3\times10^1$ & - & $2.5\times10^2$ \\
& Ne X Ly$\alpha$ & \textless$1.2\times10^2$  & \textless$1.5\times10^2$ & $8.2\times10^1$  \\
& Si XIV Ly$\alpha$ & -  & \textless$1.2\times10^2$  & $8.1\times10^1$ \\
& S XVI Ly$\alpha$ & - & \textless$8.2\times10^1$  & $5.6\times10^1$ \\
& Ar XVIII Ly$\alpha$ & - & \textless$1.1\times10^2$  & $2.0\times10^1$ \\
& Ca XX Ly$\alpha$ & - &  \textless$2.6\times10^2$ & $1.6\times10^1$ \\
\enddata
\tablecomments{
(1) Source name.
(2) Line labels of characteristic transition lines  generated in hot winds.
(3) - (4) Observed line luminosity in units of $\rm 10^{37}~erg~s^{-1}$. For lines with no prominent detection, we give 3-$\sigma$ upper limits where blue-shifts of line centers are fixed at $1.9\times10^3$ km/s for M81* and $2.6\times10^3$ for NGC\,7213).
(5) Predicted line luminosity from hot wind simulations in units of $\rm 10^{37}~erg~s^{-1}$.
}
\vspace{-3em}
\end{deluxetable}

\section{Simulated hot wind spectra in soft X-ray band}\label{appendix:hwsimu}
We follow the method described in \citetalias{2021NatAs...5..928S} to generate synthetic spectral models over 0.5-2 keV bands based on customized wind numerical simulation. For simplification, self-absorption is ignored and collisional ionization equilibrium is assumed. Intensity of optically thin thermal emission emitted by hot wind is proportional to square of wind number density, which is treated as scale-free in numerical simulation.
So the spectral model is normalized comparing to the observed spectra to determine the real wind density.
At first, we adopt the same normalization that best match 5-8 keV spectra, but it fail to adequately match the soft X-ray bands.
We find that, to achieve a legitimate description of both continuum and O VIII Ly$\alpha$ / Ne X Ly$\alpha$ line intensity in 0.5-2 keV, the normalization of hard-band free-expanding hot wind spectral models must be scaled up by a factor of $\sim$7 for M81* and scaled down by a factor of $\sim0.7$ for NGC\,7213. The comparison has been illustrated in Figure \ref{fig:hwModSimu}.
The results imply that free expanding of hot wind alone cannot simultaneously describe detected outflowing components in soft and hard X-ray bands.

\setcounter{figure}{0}
\begin{figure*}[!hpbt]
\centering
\includegraphics[width=0.7\linewidth]{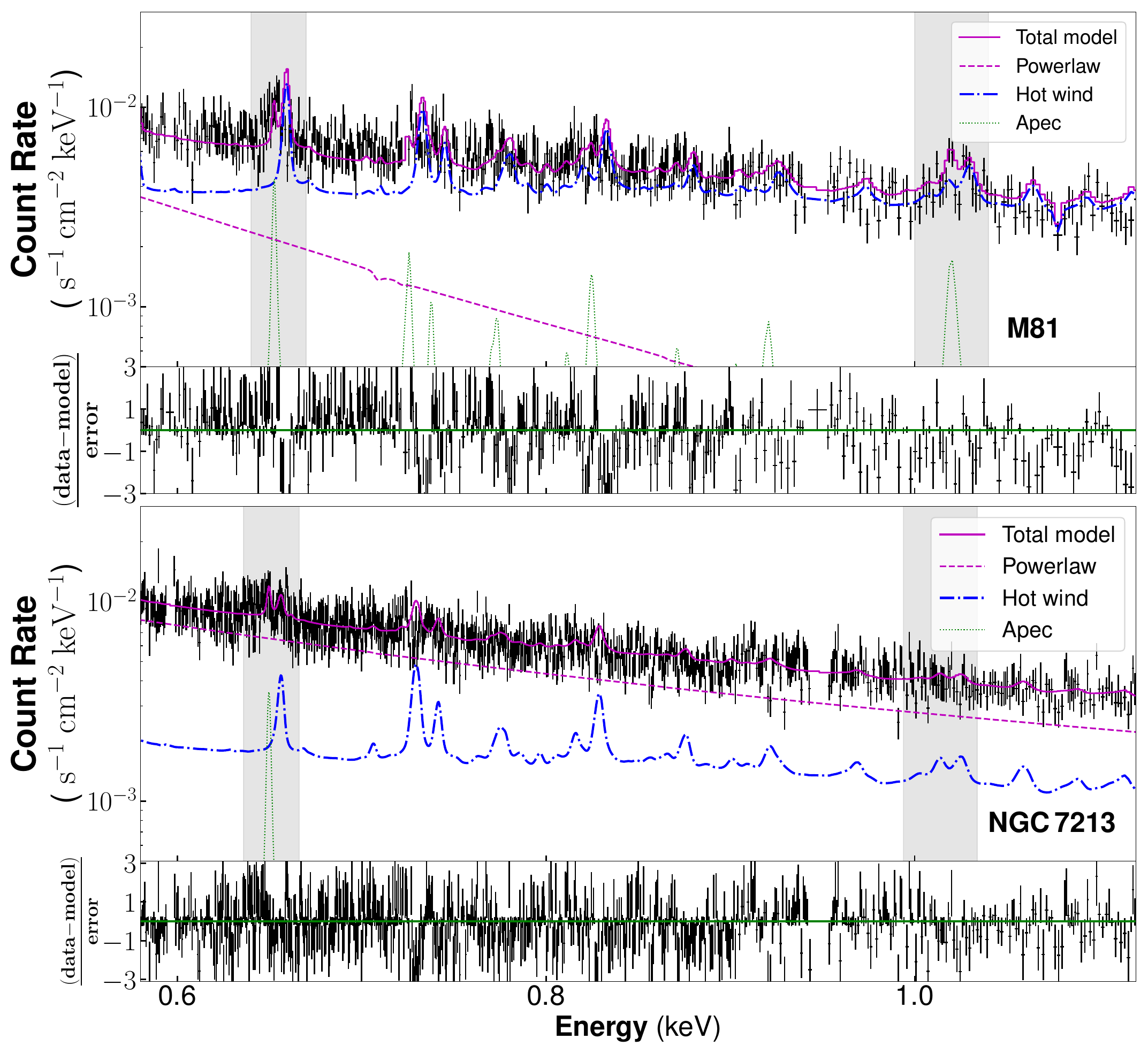}
\caption{
Comparison between synthetic hot wind spectral models from numerical simulations in \citepalias{2021NatAs...5..928S, 2022ApJ...926..209S} and observed spectra in 0.5-2 keV band. Normalization of the models has been accordingly adjusted.
The legends are the same as Fig.\ref{fig:BLsearch} (b) \& (d).
\label{fig:hwModSimu}
}
\end{figure*}

\section{Shocked clumpy circumnuclear medium}\label{appendix:clump}

We did 2-D hydrodynamical simulations describing the interaction between hot winds and clumpy warm circumnuclear medium constructed based on {\sc FLASH} v4.6 code \citep{2000ApJS..131..273F}.
Radiative cooling has been included. The applied cooling function has been calculated assuming collisional and photoionization equilibrium confined to $10^4$-$10^8$ K same as described in \citet{2019MNRAS.487..737J}. And the 'exact' cooling algorithm \citep{2009ApJS..181..391T} has been implemented to preserve the accuracy of calculations and avoid very short cooling time limiting the simulation time-step.

The simulations operate in a 4 pc$\times$4 pc box with a resolution 4096$\times$4096 in the x and y-axis direction.
Hot winds are injected at the inner boundary sphere with radius of 0.1 pc around the central point and freely expand and encounter with clumpy cool clouds distributed in the circumnuclear area.
The injected wind properties are taken from previous spectral analysis in \citepalias{2021NatAs...5..928S, 2022ApJ...926..209S} and also described in Sec. \ref{sec:diss}. 
Following the radial profile of wind density and temperature Eq. (\ref{eq:denWm81}) - Eq. (\ref{eq:tWngc7213}) and assuming the winds are freely expanding from the truncated radius to the inner boundary 0.1 pc,
for M81*, the injected wind velocity is set to $4000$ km/s, with temperature of $2.2\times10^7\rm~K$ and density of $1.8\times10^3\rm~cm^{-3}$.
For NGC\,7213 whose accretion rate is almost 30 times higher, wind velocity is set to be $7000$ km/s with temperature of $5.5\times10^7\rm~K$ and density of $1.6\times10^4\rm~cm^{-3}$.
Detailed simulation initial setups can be referred to Table \ref{tab:initsimu}.
With these setups, the total mass outflow rate of injected hot winds at 0.1 pc in the two cases are 0.002 and 0.07 $\rm M_{\odot}~yr^{-1}$, consistent with that derived from X-ray observations.
And the outer boundaries of simulation boxes are set to outflow conditions.
Spherical clumpy warm ionized clouds are randomly generated and distributed within a given radius as a mimic of the clumpy gas environment. To simplify the models, each clumpy cloud has a radius of 0.1 pc with density $2\times10^4\rm~cm^{-3}$ and temperature $1\times10^4\rm K$. These dense warm ionized clumps are buried in diffuse $10^7\rm~K$ background gas with density $20\rm~cm^{-3}$ to ensure balance in pressure.

We have tested to implement clumps with variant area-filling factor $A_{\rm f}$ in our 2-D simulation, that is, to generate different numbers of clouds within different ranges around the central black hole. 
By placing 10 clouds within 0.5 pc around M81* ($A_{\rm f,M81}=0.4$) and 60 clouds within 2 pc around NGC\,7213 ($A_{\rm f,NGC7213}=0.15$), after the shock front passing through,
the emissivity-weighted temperature and radial velocity of the shocked medium are consistent with the observed value, as shown in the T-$v_r$ phase plot in Fig. \ref{fig:flash_m81} and Fig. \ref{fig:flash_n7213}.
The density distribution of initial setups, the emissivity-weighted temperature distribution together with the velocity field and the 0.5-8 keV X-ray emissivity map of the shocked medium are also illustrated in Fig. \ref{fig:flash_m81} and Fig. \ref{fig:flash_n7213}. The 0.5-8 keV X-ray emissivity are calculated assuming thermal emission of gas in collisional ionization equilibrium based on ATOMDB. 
The 2-dimension area-filling factor of these circumnuclear clumps can be converted to volume filling factor $V_{\rm f}=A_{\rm f}^{3/2}$ in 3-dimension. And the total mass of the required warm ionized gas can then be determined as $\sim90\rm~M_{\odot}$ and $\sim1350\rm~M_{\odot}$ respectively for the two sources.
The corresponding total 0.5-8 keV X-ray luminosity from the shocked medium would be $\sim3\times10^{39}\rm~erg/s$ and $2.7\times10^{41}\rm~erg/s$, calculated from the emissivity map and normalized according to volume filling factor, larger enough to account for the radiation from the blue-shifted plasma components detected in RGS spectra.
Clumpy medium can achieve large covering factor and similar X-ray intensity with smaller volume filling factor, which reduces the required amount of cool gas.
We assign $f_{\rm cloud}$ and $f_{\rm wind}$ of each computing grid are recorded to trace the fraction of the clumpy cool clouds and hot wind components.

From the simulation snapshot t=761 yr of M81* in Fig. \ref{fig:flash_m81}, 
those brightest yellow dot-like structures in the emissivity map with the strongest X-ray emission locate roughly at the windward side of those clumpy clouds and coincide with the dark dots manifesting gas temperature between $10^6$ K and $10^7$ K.
This can be more clearly addressed in the T-$v_{\rm r}$ phase plot.
There are two aggregated bright structures and a narrow stripe with prominent X-ray emission in the phase plot.
After checking the wind fraction, the bright narrow stripe-like structure with almost constant radial velocity and decreasing temperature traces the freely expanding hot winds before encountering any circumnuclear clumpy clouds.
The brightest part at the left bottom exhibits large $f_{\rm cloud}$ thus represents shocked medium. It covers the parametric space where the observed blue-shifted gas from soft X-ray spectra resides. 
The estimated emissivity-weighted average radial gas velocity in the simulation snapshot is $\sim1700\rm~km~s^{-1}$. The clumpy nature of circumnuclear medium may also enhance the anisotropy of velocity in different directions of the large opening angle winds.
Apart from the brightest structure, the discernible diffuse filamentary structure on the top right has large $f_{\rm wind}$ and corresponds to shocked wind. Gas in the artifact outer shock shells between hot winds and manually set background gas shown in the X-ray emissivity map also resides in this part.
Nevertheless, the shocked clumpy circumnuclear medium dominates the predicted X-ray emission.
The case is similar for the snapshot t=1014 yr of NGC\,7213.
Relics of shocked clumpy clouds survive the compression and destruction by passing hot wind, cool down and maintain temperature of $\sim10^7$ K, 
exhibiting the strongest X-ray emission represented by the bright yellow dots in X-ray emissivity map. 
The best-fit temperature and velocity of observed blue-shifted gas also coincide with the structure with prominent X-ray emissivity dominated by shocked medium at left-bottom in the T-$v_{\rm r}$ phase plot.

\setcounter{figure}{0}
\begin{figure*}[hbtp!]
\centering
\includegraphics[width=0.49\linewidth]{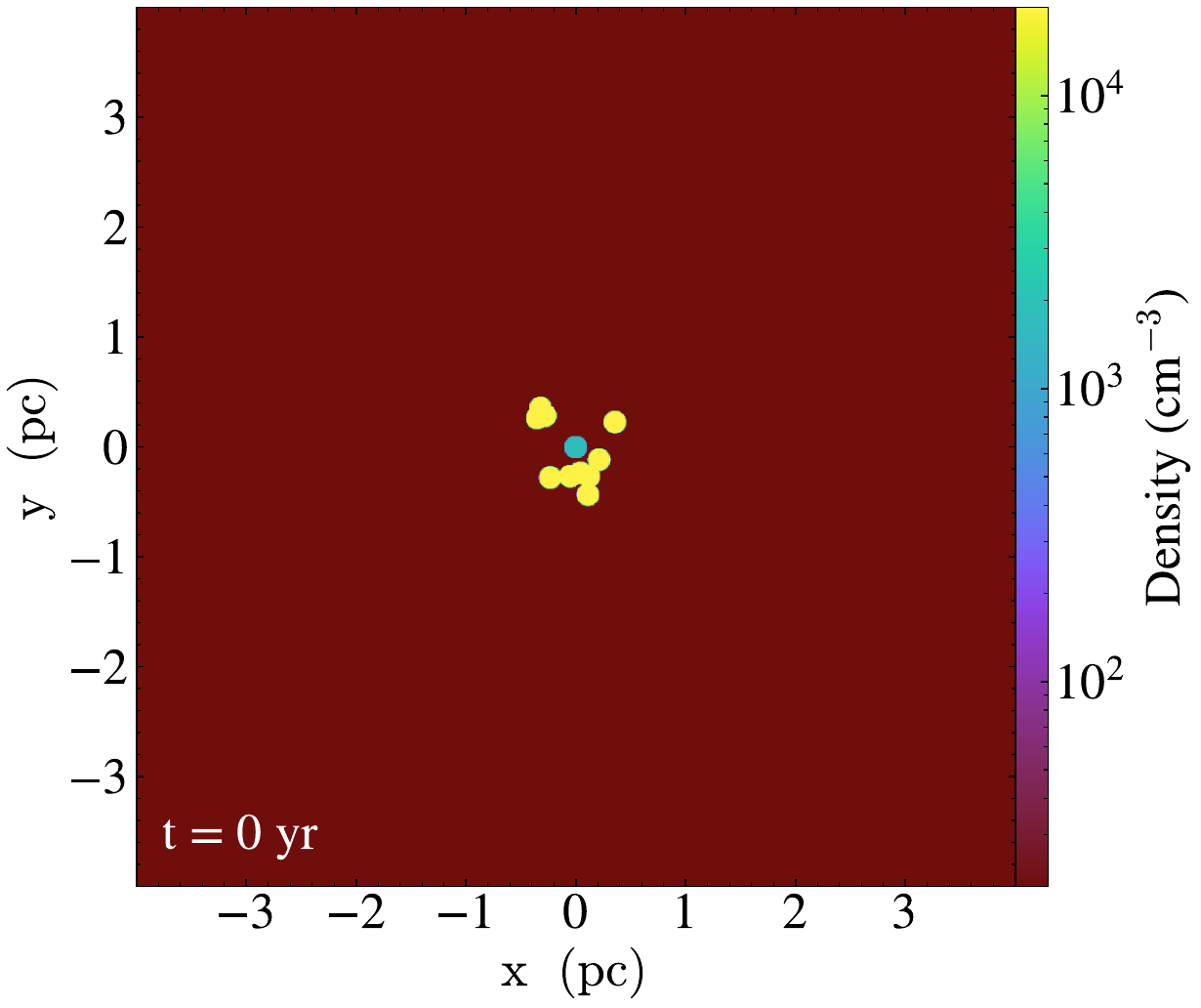}\includegraphics[width=0.49\linewidth]{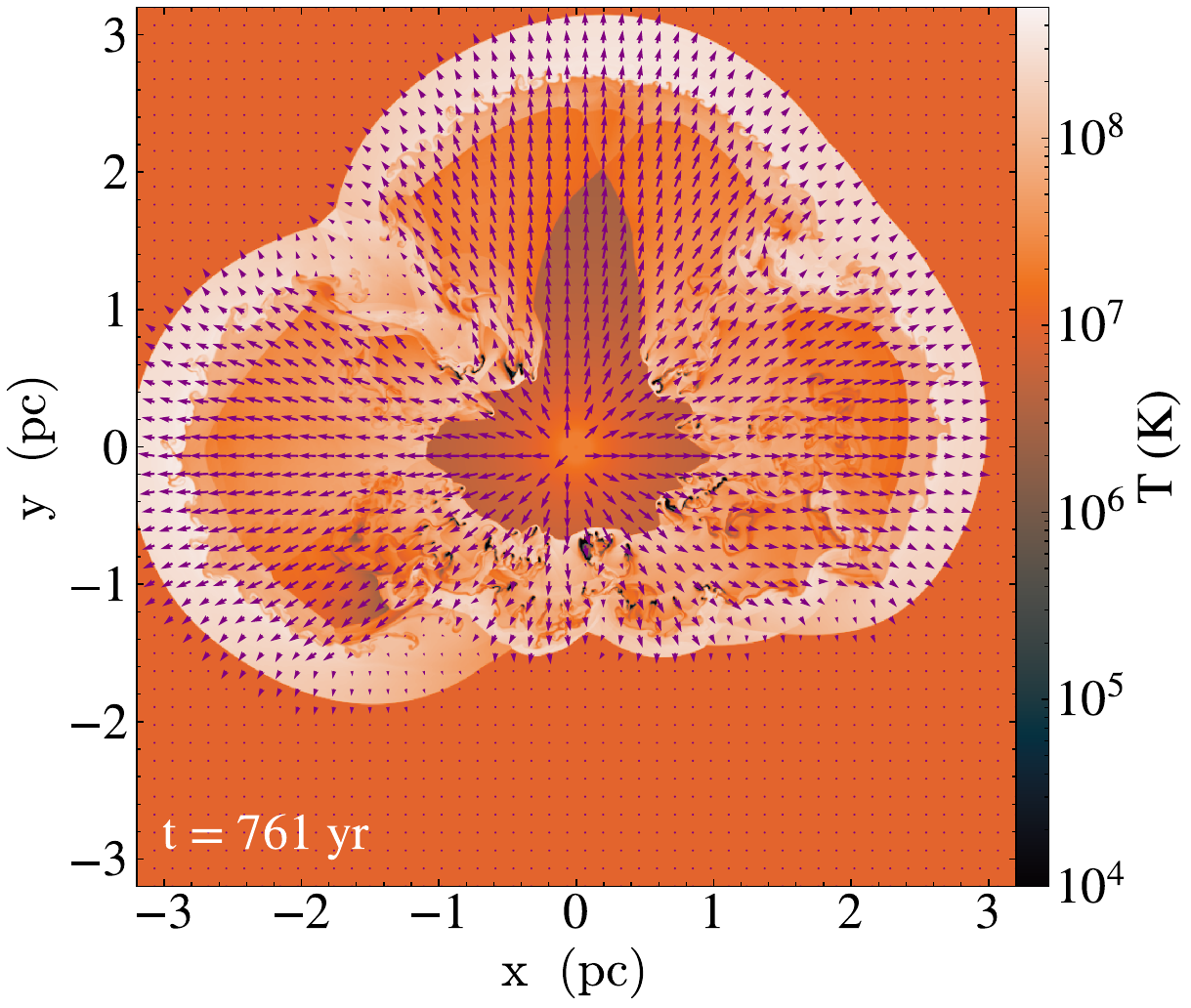}
\includegraphics[width=0.49\linewidth]{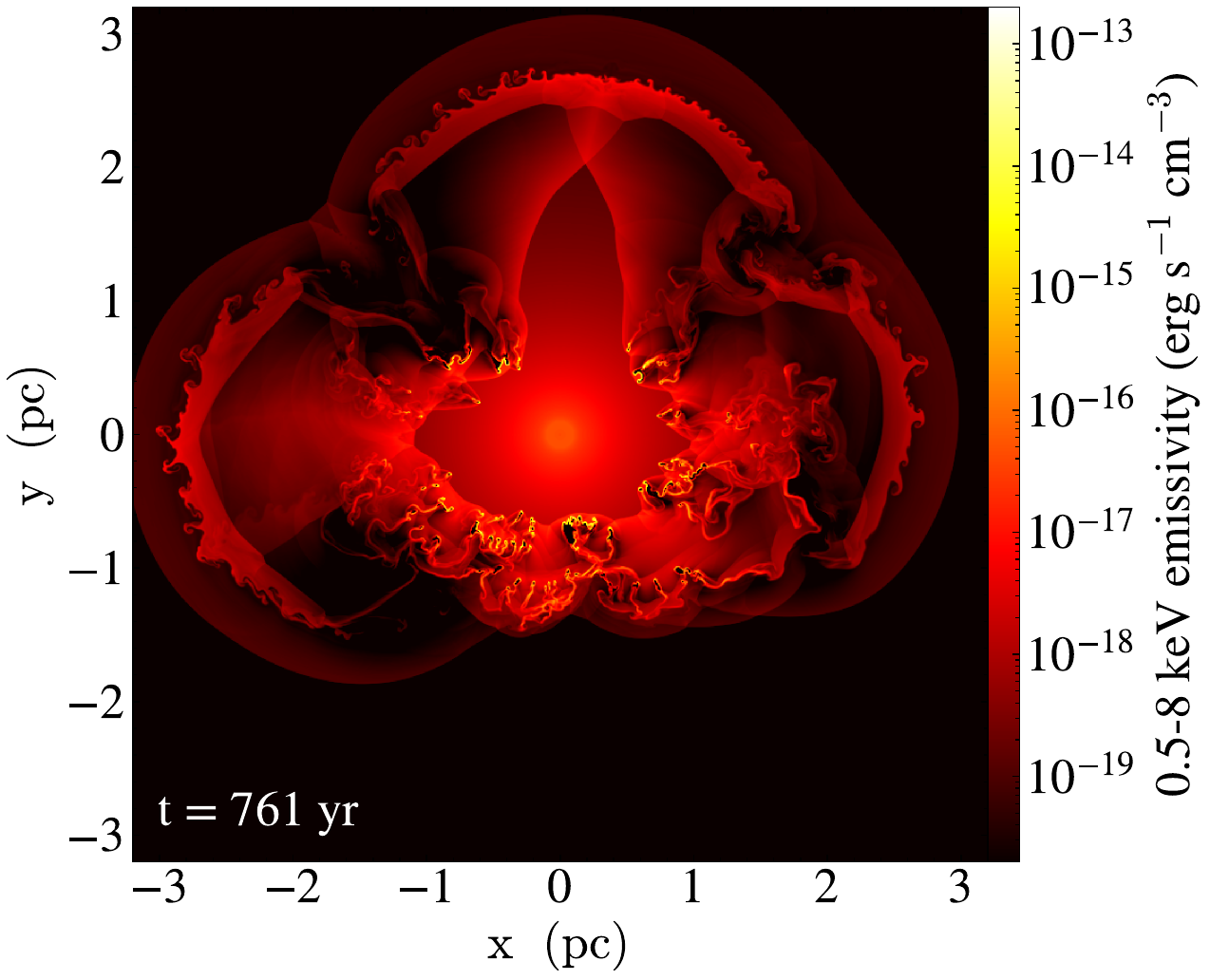}\includegraphics[width=0.49\linewidth]{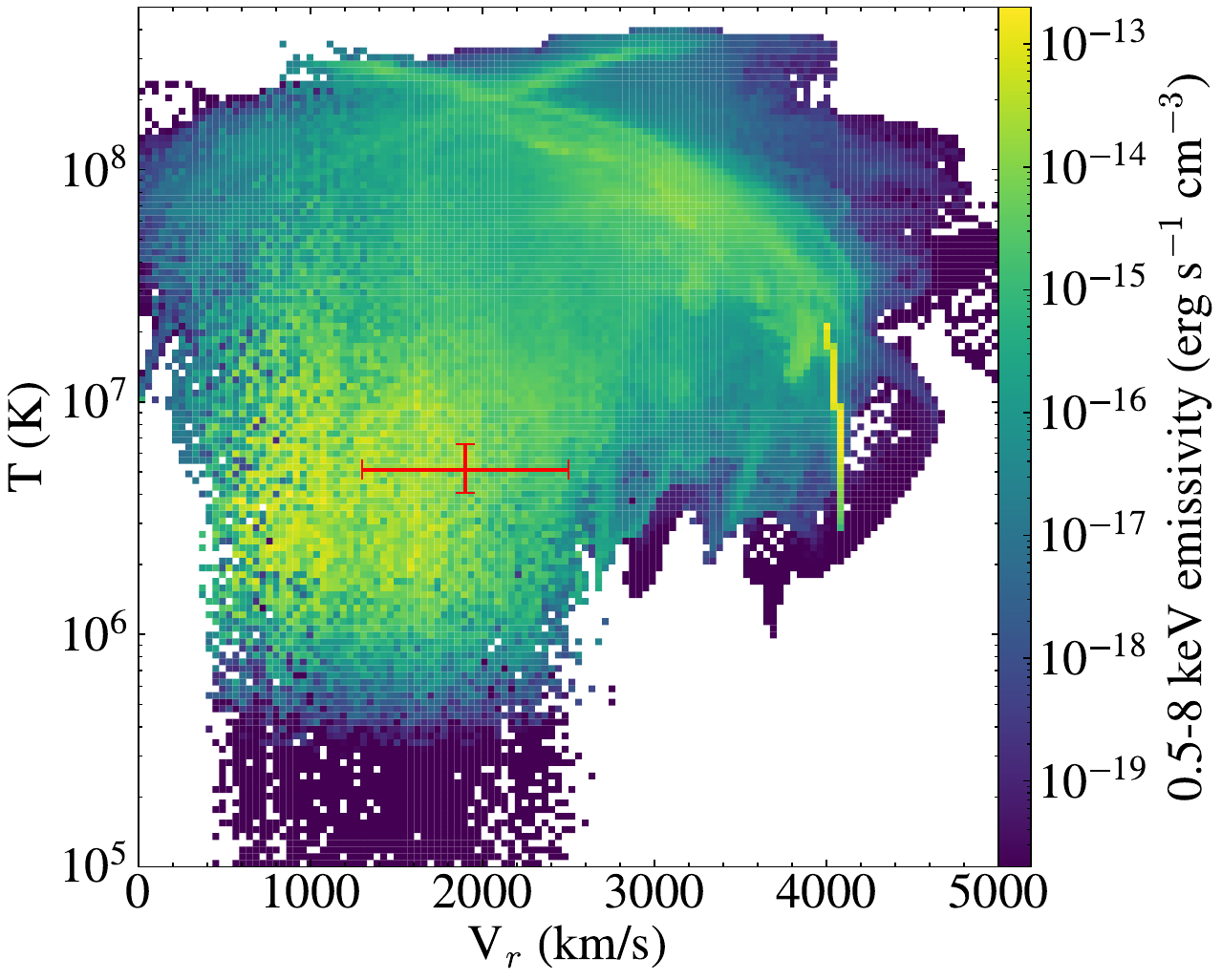}
\caption{Simulation results of circumnuclear warm ionized medium shocked by hot winds on pc-scale for M81* at snapshot t=761 yr.
{\it Top left panel}: The initial density setup. Wind is injected from a spherical with radius 0.1 pc surrounding the black hole centered on (0,0), and 10 $10^4$-K clouds are randomly distributed within 0.5 pc.
{\it Top right panel}: Temperature distribution overlapped with the velocity field.
{\it Bottom left panel}: 0.5-8 keV X-ray emissivity map.
{\it Bottom right panel}: Phase plot of gas temperature and radial velocity weighted by 0.5-8 keV X-ray emissivity. The red cross with 90\% error bars marks the best fit gas temperature and blue-shifted velocity from observed spectrum.
\label{fig:flash_m81}
}
\end{figure*}

\setcounter{table}{0}
\begin{deluxetable*}{c|ccccccccccc}[!hbtp]
\tabletypesize{\footnotesize}
\tablecaption{Initial simulation setups for wind-circumnuclear medium interaction\label{tab:initsimu}}
\tablehead{
\colhead{Source} & \colhead{code} & \colhead{$V_{\rm f}$} & \colhead{$T_{\rm in,wind}$} &
\colhead{$n_{\rm in,wind}$} & \colhead{$v_{\rm in,wind}$} & \colhead{$n_{\rm CNM}$} & \colhead{$T_{\rm CNM}$} & \colhead{$r_{\rm in,CNM}$} & \colhead{$r_{\rm out,CNM}$} & \colhead{$N_{\rm cld}$} & \colhead{$M_{\rm CNM}$} \\
\colhead{(1)} & \colhead{(2)} & \colhead{(3)} & \colhead{(4)} & \colhead{(5)} & \colhead{(6)} & \colhead{(7)} & \colhead{(8)} & \colhead{(9)} & \colhead{(10)} & \colhead{(11)} & \colhead{(12)}
}
\startdata
\multirow{2}{*}{M81*} & {\sc PLUTO} & 1 & $2.17\times10^7$ & $7.2\times10^3$ & 2800 & $2\times10^4$ & $5\times10^{3}$ & 0.1 & 1 & - & $2\times10^3$ \\
& {\sc FLASH} & 0.25 & $2.17\times10^7$ & $1.8\times10^3$ & 4000 & $2\times10^4$ & $1\times10^3$ & 0.1 & 0.5 & 10 & 90 \\
\hline
\multirow{2}{*}{NGC\,7213} & {\sc PLUTO} & 1 & $5.49\times10^7$ & $1.6\times10^4$ & 7000 & $2\times10^5$ & $5\times10^3$ & 0.1 & 1 & - & $2\times10^4$ \\
& {\sc FLASH} & 0.058 & $5.49\times10^7$ & $1.6\times10^4$ & 7000 & $2\times10^4$ & $1\times10^3$ & 0.1 & 2 & 60 & 1350\\
\enddata
\tablecomments{
(1) Source name.
(2) The adopted simulation code.
(3) Volume filling factor of circumnuclear medium. $V_{\rm f}=1$ for uniformly distributed scenario. For clumpy cloud scenario, it is derived from area filling factor $V_{\rm f}=A_{\rm f}^{3/2}$.
(4) - (6) Temperature (in units of K), density (in units of $\rm cm^{-3}$) and velocity of hot winds injected at 0.1 pc away from the central SMBH. 
(7) - (8) Temperature (in units of K) and density (in units of $\rm cm^{-3}$) of the circumnuclear medium.
(9) - (10) The inner and outer boundary of circumnuclear medium in units of pc. For clumpy cloud scenario, warm ionized gas clouds are randomly distributed between the two radius.
(11) Number of warm ionized gas clouds.
(12) Total mass of the circumnuclear medium set in the simulations, in units of $\rm M_{\odot}$.
}
\vspace{-4em}
\end{deluxetable*}

\begin{figure*}
\centering
\includegraphics[width=0.49\linewidth]{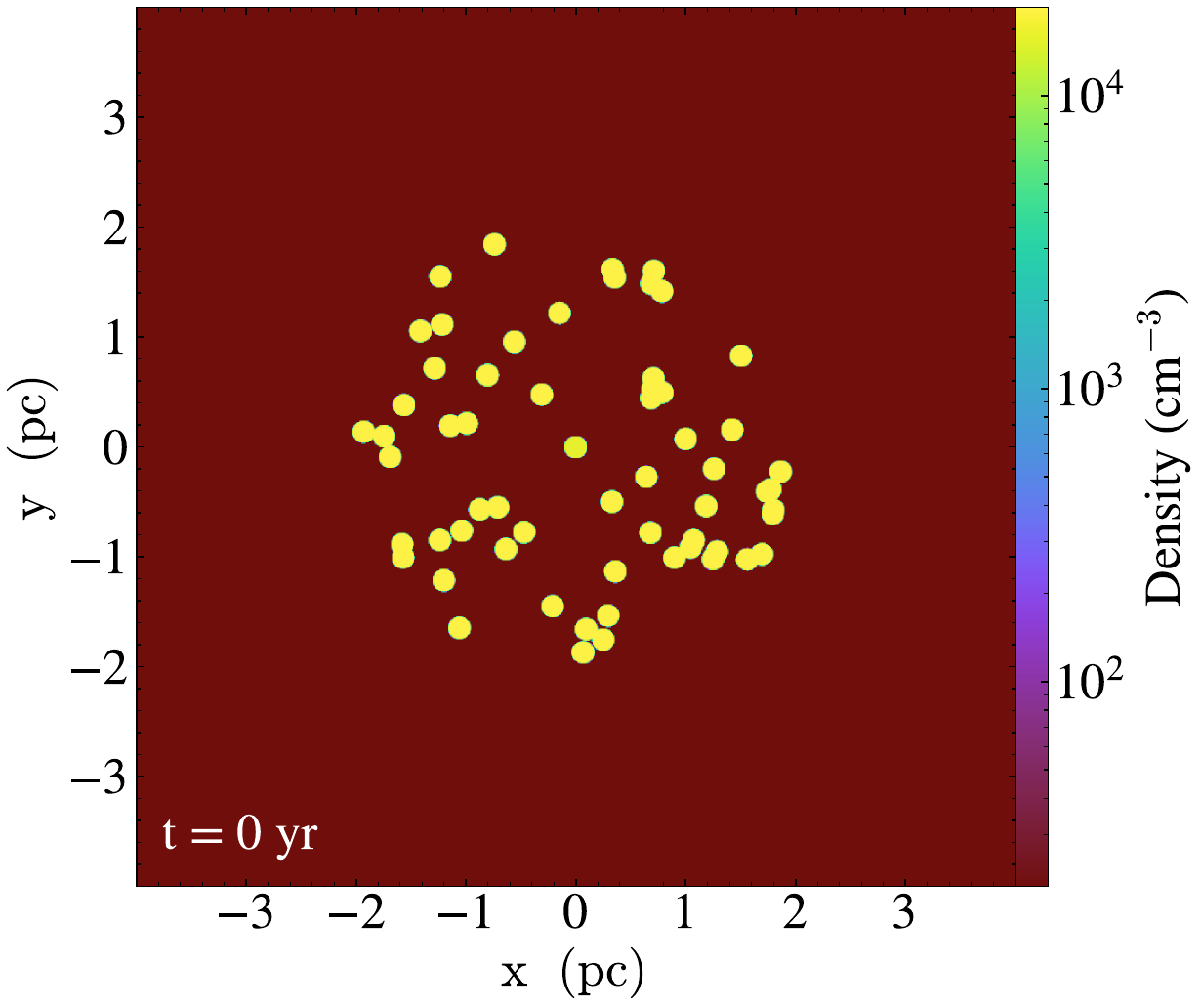}\includegraphics[width=0.49\linewidth]{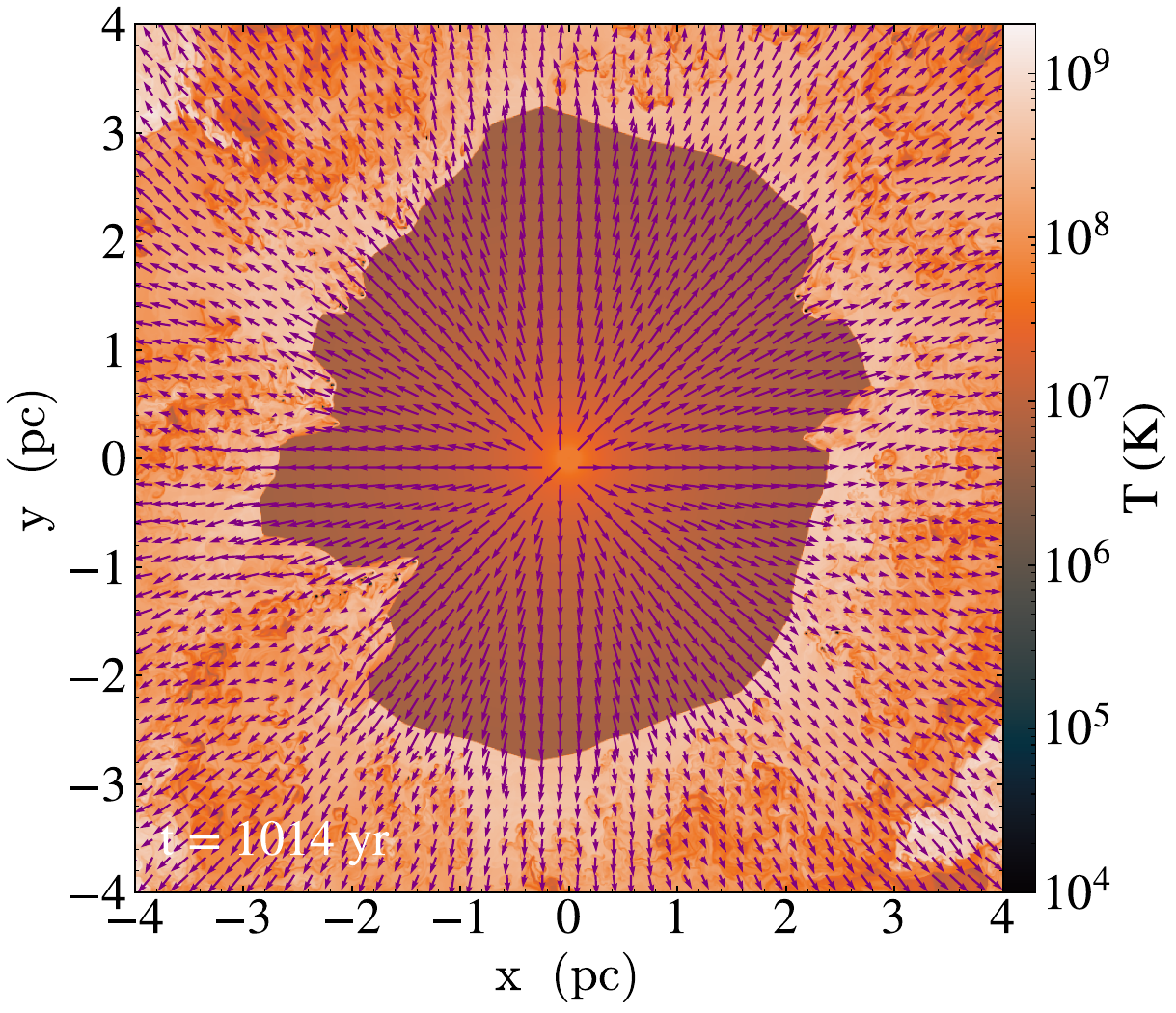}
\includegraphics[width=0.49\linewidth]{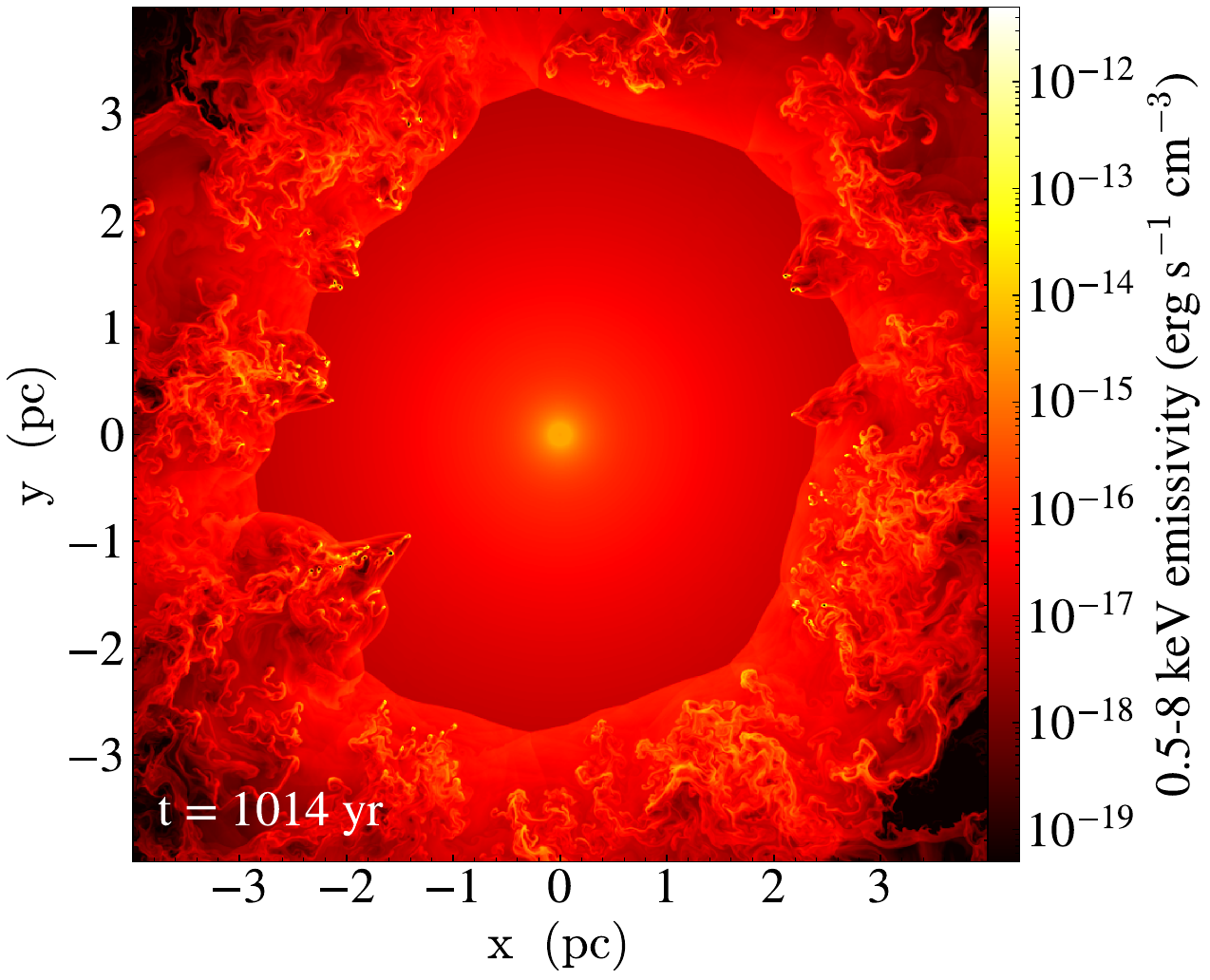}\includegraphics[width=0.49\linewidth]{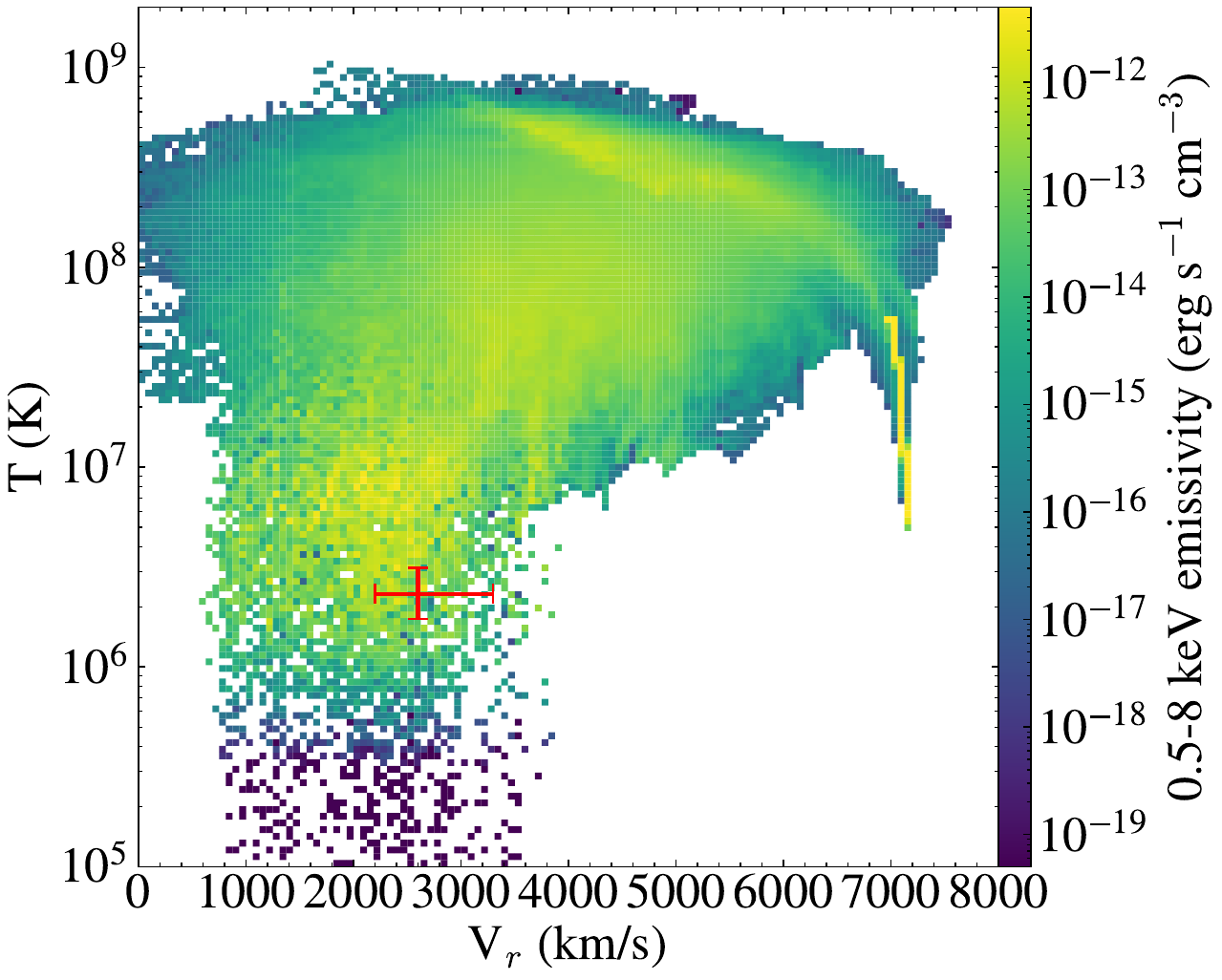}
\caption{Simulation results of circumnuclear warm ionized medium shocked by hot winds on pc-scale for NGC\,7213 at snapshot t=1014 yr. The captions are the same as Fig. \ref{fig:flash_m81}
\label{fig:flash_n7213}
}
\end{figure*}


\clearpage
\bibliography{sample631}{}

\begin{thebibliography}{}
\expandafter\ifx\csname natexlab\endcsname\relax\def\natexlab#1{#1}\fi
\providecommand{\url}[1]{\href{#1}{#1}}
\providecommand{\dodoi}[1]{doi:~\href{http://doi.org/#1}{\nolinkurl{#1}}}
\providecommand{\doeprint}[1]{\href{http://ascl.net/#1}{\nolinkurl{http://ascl.net/#1}}}
\providecommand{\doarXiv}[1]{\href{https://arxiv.org/abs/#1}{\nolinkurl{https://arxiv.org/abs/#1}}}

\bibitem[{{Arnaud}(1996)}]{1996ASPC..101...17A}
{Arnaud}, K.~A. 1996, in Astronomical Society of the Pacific Conference Series,
  Vol. 101, Astronomical Data Analysis Software and Systems V, ed. G.~H.
  {Jacoby} \& J.~{Barnes}, 17

\bibitem[{{Asplund} {et~al.}(2009){Asplund}, {Grevesse}, {Sauval}, \&
  {Scott}}]{2009ARA&A..47..481A}
{Asplund}, M., {Grevesse}, N., {Sauval}, A.~J., \& {Scott}, P. 2009, \araa, 47,
  481, \dodoi{10.1146/annurev.astro.46.060407.145222}

\bibitem[{{Astropy Collaboration} {et~al.}(2013){Astropy Collaboration},
  {Robitaille}, {Tollerud}, {Greenfield}, {Droettboom}, {Bray}, {Aldcroft},
  {Davis}, {Ginsburg}, {Price-Whelan}, {Kerzendorf}, {Conley}, {Crighton},
  {Barbary}, {Muna}, {Ferguson}, {Grollier}, {Parikh}, {Nair}, {Unther},
  {Deil}, {Woillez}, {Conseil}, {Kramer}, {Turner}, {Singer}, {Fox}, {Weaver},
  {Zabalza}, {Edwards}, {Azalee Bostroem}, {Burke}, {Casey}, {Crawford},
  {Dencheva}, {Ely}, {Jenness}, {Labrie}, {Lim}, {Pierfederici}, {Pontzen},
  {Ptak}, {Refsdal}, {Servillat}, \& {Streicher}}]{2013A&A...558A..33A}
{Astropy Collaboration}, {Robitaille}, T.~P., {Tollerud}, E.~J., {et~al.} 2013,
  \aap, 558, A33, \dodoi{10.1051/0004-6361/201322068}

\bibitem[{{Astropy Collaboration} {et~al.}(2018){Astropy Collaboration},
  {Price-Whelan}, {Sip{\H{o}}cz}, {G{\"u}nther}, {Lim}, {Crawford}, {Conseil},
  {Shupe}, {Craig}, {Dencheva}, {Ginsburg}, {VanderPlas}, {Bradley},
  {P{\'e}rez-Su{\'a}rez}, {de Val-Borro}, {Aldcroft}, {Cruz}, {Robitaille},
  {Tollerud}, {Ardelean}, {Babej}, {Bach}, {Bachetti}, {Bakanov}, {Bamford},
  {Barentsen}, {Barmby}, {Baumbach}, {Berry}, {Biscani}, {Boquien}, {Bostroem},
  {Bouma}, {Brammer}, {Bray}, {Breytenbach}, {Buddelmeijer}, {Burke},
  {Calderone}, {Cano Rodr{\'\i}guez}, {Cara}, {Cardoso}, {Cheedella}, {Copin},
  {Corrales}, {Crichton}, {D'Avella}, {Deil}, {Depagne}, {Dietrich}, {Donath},
  {Droettboom}, {Earl}, {Erben}, {Fabbro}, {Ferreira}, {Finethy}, {Fox},
  {Garrison}, {Gibbons}, {Goldstein}, {Gommers}, {Greco}, {Greenfield},
  {Groener}, {Grollier}, {Hagen}, {Hirst}, {Homeier}, {Horton}, {Hosseinzadeh},
  {Hu}, {Hunkeler}, {Ivezi{\'c}}, {Jain}, {Jenness}, {Kanarek}, {Kendrew},
  {Kern}, {Kerzendorf}, {Khvalko}, {King}, {Kirkby}, {Kulkarni}, {Kumar},
  {Lee}, {Lenz}, {Littlefair}, {Ma}, {Macleod}, {Mastropietro}, {McCully},
  {Montagnac}, {Morris}, {Mueller}, {Mumford}, {Muna}, {Murphy}, {Nelson},
  {Nguyen}, {Ninan}, {N{\"o}the}, {Ogaz}, {Oh}, {Parejko}, {Parley}, {Pascual},
  {Patil}, {Patil}, {Plunkett}, {Prochaska}, {Rastogi}, {Reddy Janga},
  {Sabater}, {Sakurikar}, {Seifert}, {Sherbert}, {Sherwood-Taylor}, {Shih},
  {Sick}, {Silbiger}, {Singanamalla}, {Singer}, {Sladen}, {Sooley},
  {Sornarajah}, {Streicher}, {Teuben}, {Thomas}, {Tremblay}, {Turner},
  {Terr{\'o}n}, {van Kerkwijk}, {de la Vega}, {Watkins}, {Weaver}, {Whitmore},
  {Woillez}, {Zabalza}, \& {Astropy Contributors}}]{2018AJ....156..123A}
{Astropy Collaboration}, {Price-Whelan}, A.~M., {Sip{\H{o}}cz}, B.~M., {et~al.}
  2018, \aj, 156, 123, \dodoi{10.3847/1538-3881/aabc4f}

\bibitem[{{Barai} {et~al.}(2012){Barai}, {Proga}, \&
  {Nagamine}}]{2012MNRAS.424..728B}
{Barai}, P., {Proga}, D., \& {Nagamine}, K. 2012, \mnras, 424, 728,
  \dodoi{10.1111/j.1365-2966.2012.21260.x}

\bibitem[{{Bianchi} {et~al.}(2008){Bianchi}, {La Franca}, {Matt}, {Guainazzi},
  {Jimenez Bail{\'o}n}, {Longinotti}, {Nicastro}, \&
  {Pentericci}}]{2008MNRAS.389L..52B}
{Bianchi}, S., {La Franca}, F., {Matt}, G., {et~al.} 2008, \mnras, 389, L52,
  \dodoi{10.1111/j.1745-3933.2008.00521.x}

\bibitem[{{Bianchi} {et~al.}(2003){Bianchi}, {Matt}, {Balestra}, \&
  {Perola}}]{2003A&A...407L..21B}
{Bianchi}, S., {Matt}, G., {Balestra}, I., \& {Perola}, G.~C. 2003, \aap, 407,
  L21, \dodoi{10.1051/0004-6361:20031054}

\bibitem[{{Cui} \& {Yuan}(2020)}]{2020ApJ...890...81C}
{Cui}, C., \& {Yuan}, F. 2020, \apj, 890, 81, \dodoi{10.3847/1538-4357/ab6e6f}

\bibitem[{{Cui} {et~al.}(2020){Cui}, {Yuan}, \& {Li}}]{2020ApJ...890...80C}
{Cui}, C., {Yuan}, F., \& {Li}, B. 2020, \apj, 890, 80,
  \dodoi{10.3847/1538-4357/ab6e6e}

\bibitem[{{Devereux}(2019)}]{2019MNRAS.488.1199D}
{Devereux}, N. 2019, \mnras, 488, 1199, \dodoi{10.1093/mnras/stz1761}

\bibitem[{{Emmanoulopoulos} {et~al.}(2012){Emmanoulopoulos}, {Papadakis},
  {McHardy}, {Ar{\'e}valo}, {Calvelo}, \& {Uttley}}]{2012MNRAS.424.1327E}
{Emmanoulopoulos}, D., {Papadakis}, I.~E., {McHardy}, I.~M., {et~al.} 2012,
  \mnras, 424, 1327, \dodoi{10.1111/j.1365-2966.2012.21316.x}

\bibitem[{{Emmanoulopoulos} {et~al.}(2013){Emmanoulopoulos}, {Papadakis},
  {Nicastro}, \& {McHardy}}]{2013MNRAS.429.3439E}
{Emmanoulopoulos}, D., {Papadakis}, I.~E., {Nicastro}, F., \& {McHardy}, I.~M.
  2013, \mnras, 429, 3439, \dodoi{10.1093/mnras/sts610}

\bibitem[{{Ferland} {et~al.}(2017){Ferland}, {Chatzikos}, {Guzm{\'a}n},
  {Lykins}, {van Hoof}, {Williams}, {Abel}, {Badnell}, {Keenan}, {Porter}, \&
  {Stancil}}]{2017RMxAA..53..385F}
{Ferland}, G.~J., {Chatzikos}, M., {Guzm{\'a}n}, F., {et~al.} 2017, \rmxaa, 53,
  385, \dodoi{10.48550/arXiv.1705.10877}

\bibitem[{{Fruscione} {et~al.}(2006){Fruscione}, {McDowell}, {Allen},
  {Brickhouse}, {Burke}, {Davis}, {Durham}, {Elvis}, {Galle}, {Harris},
  {Huenemoerder}, {Houck}, {Ishibashi}, {Karovska}, {Nicastro}, {Noble},
  {Nowak}, {Primini}, {Siemiginowska}, {Smith}, \&
  {Wise}}]{2006SPIE.6270E..1VF}
{Fruscione}, A., {McDowell}, J.~C., {Allen}, G.~E., {et~al.} 2006, in Society
  of Photo-Optical Instrumentation Engineers (SPIE) Conference Series, Vol.
  6270, Society of Photo-Optical Instrumentation Engineers (SPIE) Conference
  Series, ed. D.~R. {Silva} \& R.~E. {Doxsey}, 62701V,
  \dodoi{10.1117/12.671760}

\bibitem[{{Fryxell} {et~al.}(2000){Fryxell}, {Olson}, {Ricker}, {Timmes},
  {Zingale}, {Lamb}, {MacNeice}, {Rosner}, {Truran}, \&
  {Tufo}}]{2000ApJS..131..273F}
{Fryxell}, B., {Olson}, K., {Ricker}, P., {et~al.} 2000, \apjs, 131, 273,
  \dodoi{10.1086/317361}

\bibitem[{{Gabriel} {et~al.}(2004){Gabriel}, {Denby}, {Fyfe}, {Hoar}, {Ibarra},
  {Ojero}, {Osborne}, {Saxton}, {Lammers}, \& {Vacanti}}]{2004ASPC..314..759G}
{Gabriel}, C., {Denby}, M., {Fyfe}, D.~J., {et~al.} 2004, in Astronomical
  Society of the Pacific Conference Series, Vol. 314, Astronomical Data
  Analysis Software and Systems (ADASS) XIII, ed. F.~{Ochsenbein}, M.~G.
  {Allen}, \& D.~{Egret}, 759

\bibitem[{{Ji} {et~al.}(2019){Ji}, {Oh}, \& {Masterson}}]{2019MNRAS.487..737J}
{Ji}, S., {Oh}, S.~P., \& {Masterson}, P. 2019, \mnras, 487, 737,
  \dodoi{10.1093/mnras/stz1248}

\bibitem[{{Kormendy} \& {Ho}(2013)}]{2013ARA&A..51..511K}
{Kormendy}, J., \& {Ho}, L.~C. 2013, \araa, 51, 511,
  \dodoi{10.1146/annurev-astro-082708-101811}

\bibitem[{{Li} {et~al.}(2022){Li}, {Li}, {Garc{\'\i}a-Benito}, \&
  {Feng}}]{2022ApJ...928..111L}
{Li}, Z., {Li}, Z., {Garc{\'\i}a-Benito}, R., \& {Feng}, S. 2022, \apj, 928,
  111, \dodoi{10.3847/1538-4357/ac56d9}

\bibitem[{{Mignone} {et~al.}(2007){Mignone}, {Bodo}, {Massaglia}, {Matsakos},
  {Tesileanu}, {Zanni}, \& {Ferrari}}]{2007ApJS..170..228M}
{Mignone}, A., {Bodo}, G., {Massaglia}, S., {et~al.} 2007, \apjs, 170, 228,
  \dodoi{10.1086/513316}

\bibitem[{{Narayan} {et~al.}(2012){Narayan}, {S{\"A} dowski}, {Penna}, \&
  {Kulkarni}}]{2012MNRAS.426.3241N}
{Narayan}, R., {S{\"A} dowski}, A., {Penna}, R.~F., \& {Kulkarni}, A.~K. 2012,
  \mnras, 426, 3241, \dodoi{10.1111/j.1365-2966.2012.22002.x}

\bibitem[{{Nemmen} {et~al.}(2014){Nemmen}, {Storchi-Bergmann}, \&
  {Eracleous}}]{2014MNRAS.438.2804N}
{Nemmen}, R.~S., {Storchi-Bergmann}, T., \& {Eracleous}, M. 2014, \mnras, 438,
  2804, \dodoi{10.1093/mnras/stt2388}

\bibitem[{{Osterbrock} \& {Ferland}(2006)}]{2006agna.book.....O}
{Osterbrock}, D.~E., \& {Ferland}, G.~J. 2006, {Astrophysics of gaseous nebulae
  and active galactic nuclei}

\bibitem[{{Ploeckinger} \& {Schaye}(2020)}]{2020MNRAS.497.4857P}
{Ploeckinger}, S., \& {Schaye}, J. 2020, \mnras, 497, 4857,
  \dodoi{10.1093/mnras/staa2172}

\bibitem[{{Rybicki} \& {Lightman}(1979)}]{1979rpa..book.....R}
{Rybicki}, G.~B., \& {Lightman}, A.~P. 1979, {Radiative processes in
  astrophysics}

\bibitem[{{Salvestrini} {et~al.}(2020){Salvestrini}, {Gruppioni}, {Pozzi},
  {Vignali}, {Giannetti}, {Paladino}, \&
  {Hatziminaoglou}}]{2020A&A...641A.151S}
{Salvestrini}, F., {Gruppioni}, C., {Pozzi}, F., {et~al.} 2020, \aap, 641,
  A151, \dodoi{10.1051/0004-6361/202037660}

\bibitem[{{Schimoia} {et~al.}(2017){Schimoia}, {Storchi-Bergmann}, {Winge},
  {Nemmen}, \& {Eracleous}}]{2017MNRAS.472.2170S}
{Schimoia}, J.~S., {Storchi-Bergmann}, T., {Winge}, C., {Nemmen}, R.~S., \&
  {Eracleous}, M. 2017, \mnras, 472, 2170, \dodoi{10.1093/mnras/stx2107}

\bibitem[{{Schnorr-M{\"u}ller} {et~al.}(2014){Schnorr-M{\"u}ller},
  {Storchi-Bergmann}, {Nagar}, \& {Ferrari}}]{2014MNRAS.438.3322S}
{Schnorr-M{\"u}ller}, A., {Storchi-Bergmann}, T., {Nagar}, N.~M., \& {Ferrari},
  F. 2014, \mnras, 438, 3322, \dodoi{10.1093/mnras/stt2440}

\bibitem[{{Shi} {et~al.}(2021){Shi}, {Li}, {Yuan}, \&
  {Zhu}}]{2021NatAs...5..928S}
{Shi}, F., {Li}, Z., {Yuan}, F., \& {Zhu}, B. 2021, Nature Astronomy, 5, 928,
  \dodoi{10.1038/s41550-021-01394-0}

\bibitem[{{Shi} {et~al.}(2022){Shi}, {Zhu}, {Li}, \&
  {Yuan}}]{2022ApJ...926..209S}
{Shi}, F., {Zhu}, B., {Li}, Z., \& {Yuan}, F. 2022, \apj, 926, 209,
  \dodoi{10.3847/1538-4357/ac4789}

\bibitem[{{Starling} {et~al.}(2005){Starling}, {Page}, {Branduardi-Raymont},
  {Breeveld}, {Soria}, \& {Wu}}]{2005MNRAS.356..727S}
{Starling}, R.~L.~C., {Page}, M.~J., {Branduardi-Raymont}, G., {et~al.} 2005,
  \mnras, 356, 727, \dodoi{10.1111/j.1365-2966.2004.08493.x}

\bibitem[{{Tombesi} {et~al.}(2010){Tombesi}, {Cappi}, {Reeves}, {Palumbo},
  {Yaqoob}, {Braito}, \& {Dadina}}]{2010A&A...521A..57T}
{Tombesi}, F., {Cappi}, M., {Reeves}, J.~N., {et~al.} 2010, \aap, 521, A57,
  \dodoi{10.1051/0004-6361/200913440}

\bibitem[{{Townsend}(2009)}]{2009ApJS..181..391T}
{Townsend}, R.~H.~D. 2009, \apjs, 181, 391, \dodoi{10.1088/0067-0049/181/2/391}

\bibitem[{{Yan} \& {Xie}(2018)}]{2018MNRAS.475.1190Y}
{Yan}, Z., \& {Xie}, F.-G. 2018, \mnras, 475, 1190,
  \dodoi{10.1093/mnras/stx3259}

\bibitem[{{Yang} {et~al.}(2021){Yang}, {Yuan}, {Yuan}, \&
  {White}}]{2021ApJ...914..131Y}
{Yang}, H., {Yuan}, F., {Yuan}, Y.-F., \& {White}, C.~J. 2021, \apj, 914, 131,
  \dodoi{10.3847/1538-4357/abfe63}

\bibitem[{{Yoon} {et~al.}(2019){Yoon}, {Yuan}, {Ostriker}, {Ciotti}, \&
  {Zhu}}]{2019ApJ...885...16Y}
{Yoon}, D., {Yuan}, F., {Ostriker}, J.~P., {Ciotti}, L., \& {Zhu}, B. 2019,
  \apj, 885, 16, \dodoi{10.3847/1538-4357/ab45e8}

\bibitem[{{Yu} {et~al.}(2011){Yu}, {Yuan}, \& {Ho}}]{2011ApJ...726...87Y}
{Yu}, Z., {Yuan}, F., \& {Ho}, L.~C. 2011, \apj, 726, 87,
  \dodoi{10.1088/0004-637X/726/2/87}

\bibitem[{{Yuan} {et~al.}(2012{\natexlab{a}}){Yuan}, {Bu}, \&
  {Wu}}]{2012ApJ...761..130Y}
{Yuan}, F., {Bu}, D., \& {Wu}, M. 2012{\natexlab{a}}, \apj, 761, 130,
  \dodoi{10.1088/0004-637X/761/2/130}

\bibitem[{{Yuan} {et~al.}(2015){Yuan}, {Gan}, {Narayan}, {Sadowski}, {Bu}, \&
  {Bai}}]{2015ApJ...804..101Y}
{Yuan}, F., {Gan}, Z., {Narayan}, R., {et~al.} 2015, \apj, 804, 101,
  \dodoi{10.1088/0004-637X/804/2/101}

\bibitem[{{Yuan} \& {Narayan}(2014)}]{2014ARA&A..52..529Y}
{Yuan}, F., \& {Narayan}, R. 2014, \araa, 52, 529,
  \dodoi{10.1146/annurev-astro-082812-141003}

\bibitem[{{Yuan} {et~al.}(2012{\natexlab{b}}){Yuan}, {Wu}, \&
  {Bu}}]{2012ApJ...761..129Y}
{Yuan}, F., {Wu}, M., \& {Bu}, D. 2012{\natexlab{b}}, \apj, 761, 129,
  \dodoi{10.1088/0004-637X/761/2/129}

\bibitem[{{Yuan} {et~al.}(2018){Yuan}, {Yoon}, {Li}, {Gan}, {Ho}, \&
  {Guo}}]{2018ApJ...857..121Y}
{Yuan}, F., {Yoon}, D., {Li}, Y.-P., {et~al.} 2018, \apj, 857, 121,
  \dodoi{10.3847/1538-4357/aab8f8}

\end{thebibliography}
\bibliographystyle{aasjournal}



\end{document}